%% file: no_comments.tex
\documentclass[acmsmall,nonacm]{acmart}

\usepackage{subfig}
\usepackage{xspace}
\AtBeginDocument{%
  \providecommand\BibTeX{{%
    \normalfont B\kern-0.5em{\scshape i\kern-0.25em b}\kern-0.8em\TeX}}}

\newcommand\blfootnote[1]{%
  \begingroup
  \renewcommand\thefootnote{}\footnote{#1}%
  \addtocounter{footnote}{-1}%
  \endgroup
}

\usepackage{tikz}
\newcommand*\circled[1]{\tikz[baseline=(char.base)]{
    \node[shape=circle,text=white,fill=black,inner sep=0.5pt] (char) {#1};}}

\newcommand{\one}{({\em i}\/)\xspace}
\newcommand{\two}{({\em ii}\/)\xspace}
\newcommand{\three}{({\em iii}\/)\xspace}
\newcommand{\four}{({\em iv}\/)\xspace}

\def\eg{\emph{e.g.}\xspace}
\def\etc{\emph{etc.}\xspace}
\def\ie{\emph{i.e.}\xspace}
\def\etal{\emph{et al.}\xspace}
\def\vs{\emph{vs.}\xspace}

\newcommand{\pb}[1]{\vspace{0.75ex}\noindent{\bf \em #1}\hspace*{.3em}}

\newcommand\gareth[1]{\textbf{\textcolor{red}{GT: #1}}	}
\newcommand\ar[1]{\textbf{\textcolor{blue}{AR: #1}}	}
\newcommand\ignacio[1]{\textbf{\textcolor{cyan}{IC: #1}}	}
\newcommand\edc[1]{\textbf{\textcolor{brown}{EDC: #1}}	}
\newcommand\haris[1]{\textbf{\textcolor{orange}{HBZ: #1}}	}
\newcommand\ns[1]{\textbf{\textcolor{green}{NS: #1}}	}

\renewcommand\gareth[1]{} 
\renewcommand\ar[1]{} 
\renewcommand\ignacio[1]{} 
\renewcommand\edc[1]{} 
\renewcommand\haris[1]{} 
\renewcommand\ns[1]{} 

\graphicspath{ {./images/} }

\begin{document}

\title{Toxicity in the Decentralized Web and the Potential for Model Sharing$^\dag$}

\author{Haris Bin Zia}
\affiliation{%
  \institution{Queen Mary University of London}
  \country{United Kingdom}
  }
\email{h.b.zia@qmul.ac.uk}

\author{Aravindh Raman}
\affiliation{%
  \institution{Telefonica Research}
  \country{Spain}
  }
\email{aravindh.raman@telefonica.com}

\author{Ignacio Castro}
\affiliation{%
  \institution{Queen Mary University of London}
  \country{United Kingdom}
  }
\email{i.castro@qmul.ac.uk}

\author{Ishaku Hassan Anaobi}
\affiliation{%
  \institution{Queen Mary University of London}
  \country{United Kingdom}
  }
\email{i.h.anaobi@qmul.ac.uk}

\author{Emiliano De Cristofaro}
\affiliation{%
  \institution{University College London}
  \country{United Kingdom}
  }
\email{e.decristofaro@ucl.ac.uk}

\author{Nishanth Sastry}
\affiliation{%
  \institution{University of Surrey}
  \country{United Kingdom}
  }
\email{n.sastry@surrey.ac.uk}

\author{Gareth Tyson}
\affiliation{%
  \institution{Hong Kong University of Science \& Technology}
  \country{Hong Kong}
  }
 \email{gtyson@ust.hk}

\additionalaffiliation{
  \institution{Queen Mary University of London}
  }

\renewcommand{\shortauthors}{Haris Bin Zia et al.}

\begin{abstract}

The ``Decentralised Web'' (DW) is an evolving concept, which encompasses technologies aimed at providing greater transparency and openness on the web.
The DW relies on independent servers (aka instances) that mesh together in a peer-to-peer fashion to deliver a range of services (\eg micro-blogs, image sharing, video streaming).
However, toxic content moderation in this decentralised context is challenging.
This is because there is no central entity that can define toxicity, nor a large central pool of data that can be used to build universal classifiers.
It is therefore unsurprising that there have been several high-profile cases of the DW being misused to coordinate and disseminate harmful material.
Using a dataset of 9.9M posts from 117K users on Pleroma (a popular DW microblogging service), we quantify the presence of toxic content.
We find that toxic content is prevalent and spreads rapidly between instances. 
We show that automating per-instance content moderation is challenging due to the lack of sufficient training data available and the  effort required  in labelling. 
We therefore propose and evaluate \emph{ModPair}, a model sharing system that effectively detects toxic content, gaining an average per-instance macro-F1 score 0.89.

\end{abstract}

\blfootnote{$\dag$Published in the Proceedings of the 2022 ACM International Conference on Measurement and Modeling of Computer Systems (SIGMETRICS'22). Please cite accordingly.}

\maketitle

\section{Introduction}

Social platforms such as Facebook and Twitter are amongst the most popular websites in the world. 
Despite their huge success, criticism for their role in disseminating toxic content (\eg hate speech) is mounting.
This is a thorny issue
that could pose a threat to freedom of speech~\cite{freespeech}:
the monolithic nature and lack of competition
of these platforms gives them full discretion over the activities that are allowed to take place. 
This has raised a number of regulatory concerns and  triggered widespread political debate~\cite{nytimes}.

In reaction to this, a growing movement referred to as the ``Decentralised Web'' (DW) has emerged. The goal of the DW  is to decentralize power and control away from the major centralised tech giants. As such, these projects have striven to create a more open environment that champions freedom of speech, whilst simultaneously disincentivising toxicity by giving users greater control over their own online communities. 
The DW consists of a range of platforms that offer decentralised equivalents to mainstream services.
Some of the most popular DW applications include Pleroma and Mastodon~\cite{mastodonurl}~\cite{pleromaurl} (Twitter-like micro-blogging services), Diaspora~\cite{diasporaurl} (a Facebook-like social network), and PeerTube~\cite{peertubeurl} (a YouTube-like video hosting software).
These DW platforms have several unique properties: \one~They are made up of independently operated and moderated communities located on different servers (called \emph{instances}), which any new administrator can setup; \two~They enable users, who must sign-up to specific instances, to own their data -- in fact, some users choose to fork their own instance to keep complete control of their data;
and \three~They allow users to interact locally (within instances) as well as globally (across instances) via the so-called ``Fediverse'' --- this involves instances interconnecting in a peer-to-peer fashion, allowing their users to communicate (referred to as \emph{federation}).
Through these novel features, these independent servers collaborate to offer a globally integrated platform atop of an entirely decentralised infrastructure.

Despite its novelty, this model creates interesting challenges for \emph{toxic content moderation} (\S\ref{sec:background}). 
Namely, whereas centralised services (like Twitter) have the resources to actively moderate content by hiring moderators and refining automated classifiers using large data pools, 
DW administrators have limited resources and only control the data in their own instance.
Furthermore, whereas many major services rely on third-party commercial and centralised moderation  APIs~\cite{perspective} (that tag their posts), this goes against the vision and philosophy of the DW.
Even if this were possible, an administrator can only moderate their own instance --- it is possible for content from poorly moderated instances to ``spread'' to other instances through the peer-to-peer federation links.
Of course, this all assumes that administrators even have similar definitions of ``toxic''. 
Thus, addressing the challenge of toxic content moderation has proven to be labour-intensive and impractical for many instance operators. 
A recent example of this occurred when the \url{gab.social} instance~\cite{zannettou2018gab} joined the Fediverse, rapidly spreading hate speech to other instances~\cite{gabblog}.
We argue that addressing this problem is vital for ensuring the success of DW applications.

With this in mind, we identify four critical questions:
\one~How much toxicity exists in the DW?
\two~How does toxic material spread across DW instances via federation?
\three~Is it possible for DW instances to train their own automated classification models to reduce the manual load on administrators?
And \four~How can administrators cooperate to improve content moderation and reduce their workload?

To address these questions, we present the first large-scale study of toxicity in the DW
and propose \textit{ModPair}, a collaborative approach to better enable instances to automate content moderation in a decentralised fashion. 
We do so from the point of view of one of the largest decentralised microblogging networks, \textit{Pleroma}.
We first present a measurement study of Pleroma (\S\ref{sec:measurements}), gathering data from 30 unique Pleroma instances (\S\ref{sec:dataset}), 
covering more than three years and 9.9M toots\footnote{A toot is equivalent to a tweet in Twitter.} from 117K users. 
We confirm that extensive toxic content \emph{is} present in the DW.
We identify 12.15\% of all toots as toxic. Furthermore, we show that toxic content~\emph{does} spread across Pleroma: 26 out of the 30 instances receive an average of at least 105K remote toxic toots through federation.
In fact, Pleroma instances receive more toxic content from remote instances than they generate locally. This makes it impractical for individual administrators to manually flag toxic toots at such a scale.

Driven by these observations, we explore the potential of decentralised automated content moderation in the DW (\S\ref{sec:classifiers}).
We start by assessing the ability of administrators to train local models to automatically tag local toots as toxic \vs non-toxic.
We confirm that it \emph{is} possible to build such local models, attaining an average macro-F1 score of 0.84 across all instances. 
However, we find that such models struggle to accurately tag remote content due to divergent linguistic features. 
For example, whereas a model trained using \texttt{my.dirtyhobby.xyz} data gains a macro-F1 score of 0.95, it attains an average of just 0.69 when applied to toots from other instances. This makes it difficult to rely on such models to detect toxicity in incoming toots from remote instances, without requiring administrators to tag thousands of more toots.

To improve moderation, we therefore propose \emph{ModPair} (\S\ref{sec:modpair}) --- a system for collaborative moderation where instances share partially trained models to assist each other.
This has the benefit of implicitly sharing annotations across instances, without compromising the privacy of administrators by asking them to expose their own moderation tags.  
To enhance knowledge transfer, \emph{ModPair} leverages semantic similarity of instances by pairing instances with similar topical characteristics. Specifically, each instance uses its own small set of (annotated) toxic toots to build local models. Instances then predict which remote models are most likely to perform well on their own local content, creating an ensemble to improve their own classification performance. 
We show that ModPair can correctly identify the top three best performing remote models 83\% of the time, gaining an average per-instance macro-F1 score 0.89.
Although here we focus on text-based toxic content moderation, our methodology can equally apply to other types of decentralised content classification, \eg image classification (\S\ref{sec:conclusion}).

\section{Background \& Motivation}
\label{sec:background}

\pb{What is the Fediverse?} The Fediverse is a network of  independently hosted and interconnected ``Decentralised Web'' servers (\ie instances).
In this model, no single entity operates the entire infrastructure. Instead, instances collaborate (aka ``federate'') in a peer-to-peer fashion to collectively offer various types of services (\eg microblogging, file sharing, video streaming).
This federation is performed using the W3C ActivityPub~\cite{activitypub} protocol, which allows instances to subscribe to objects provided by each other. 
The nature of these objects varies based on the specific application in question. For example, whereas Pleroma (a microblogging platform) exchanges messages, PeerTube (a video sharing platform) exchanges videos. 
This allows these server instances to form a decentralised network of content exchange.
There are around 30 open source platforms that are built with ActivityPub. Servers installing any of these software packages are interoperable and can communicate with each other irrespective of the service they provide.

\pb{What is Pleroma?} Pleroma is a DW open source \emph{microblogging} service that runs in the Fediverse
and the largest decentralised social platform (next to Mastodon~\cite{raman2019challenges}). 
Unlike centralised online social networks, anyone can run a Pleroma instance which will then operate independently under the control of its administrator. The Pleroma software is a lightweight web server written in Elixir, using the Phoenix framework with a database backed with PostgreSQL. Accordingly, administrators can deploy Pleroma on a wide range of hardware, ranging from Raspberry Pis to high-end compute servers. Both administrators and regular users can then access these instances using any web browser (via HTTP).

Each Pleroma instance has a unique domain name and users must sign up to a specific instance to gain access to the wider Pleroma network.
Often, each instance will be themed around a given topic, \eg users interested in arts might join~\texttt{imaginair.es}, whereas programming might join~\texttt{pythondevs.social}. 
Once a user has created an account on an instance, they can follow other accounts from the same instance or, alternatively, users on other instances that also host Pleroma.\footnote{In fact, they can follow users from any other Fediverse instances that use the ActivityPub protocol (\eg Mastodon, Peertube, \etc).}

In Pleroma, ``toots'' are the equivalents of ``tweets'' in Twitter and much of the functionality (\eg the ability to follow users and `like'' toots) works in a similar fashion. 
That said, there are clear differences between Pleroma and prior microblogging services: Pleroma \one~provides no ranking and recommendation algorithm instead toots are displayed chronologically;
\two~has no algorithm to recommend followees, instead new connections rely on searching an already known user through the search functions or  exploring the instances to find like-minded users;
\three~is a community-oriented platform: each instance supports specific interests or topics and users can register on the instance that is better matched to their own tastes. 

\begin{figure*}[t]
\centering
\subfloat[][]{
\includegraphics[width=0.7\columnwidth]{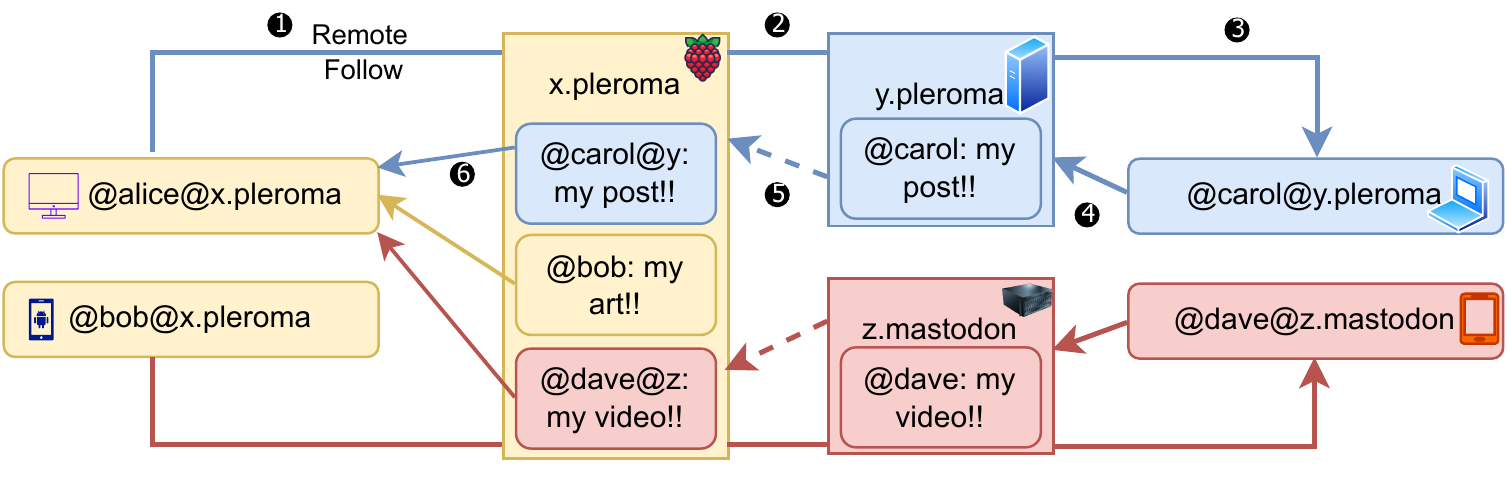} 
\label{fig:federation}
}
\subfloat[][]{
\includegraphics[width=0.28\columnwidth]{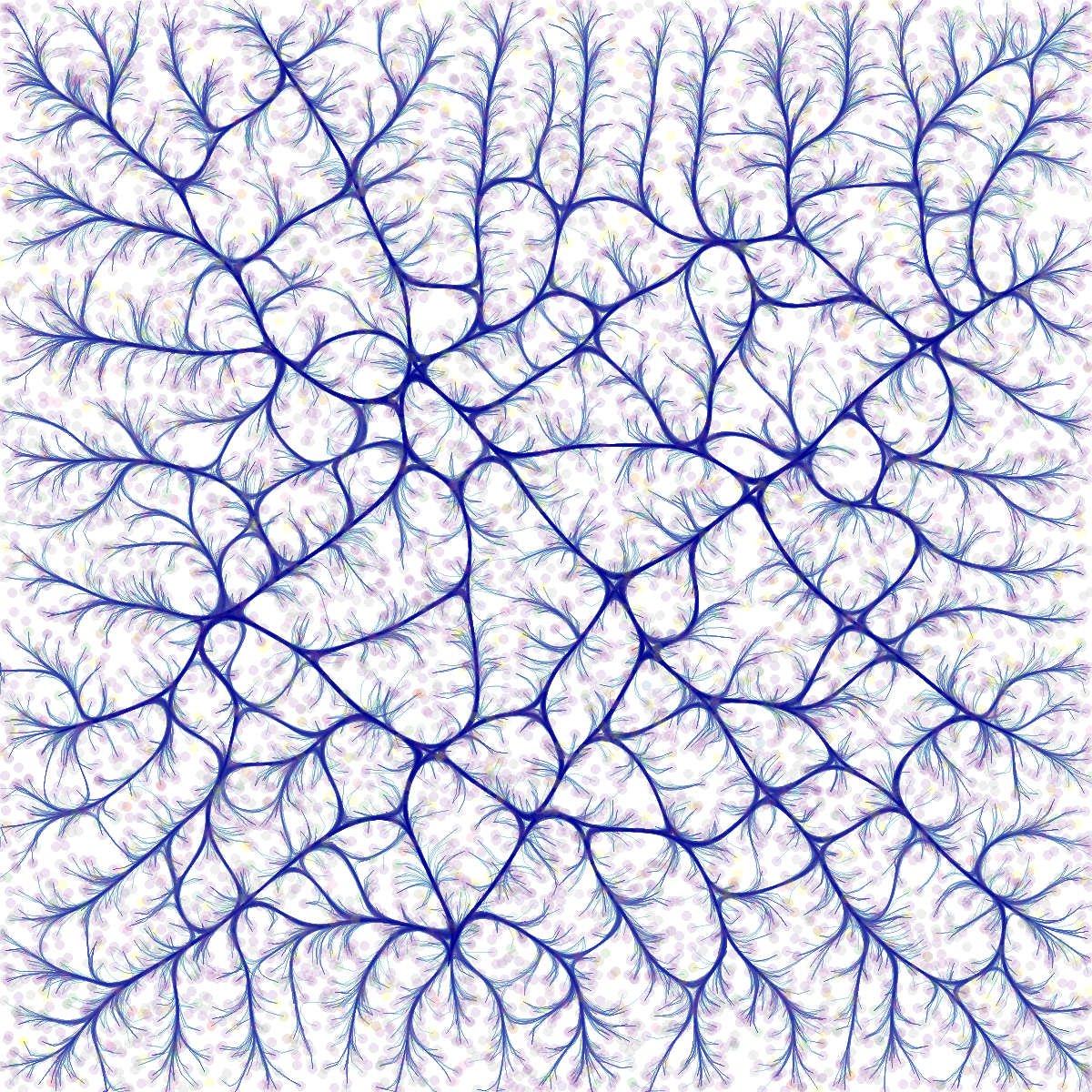}
\label{fig:pleroma30connection}
}
\caption{$(a)$ Example federation workflow. $(b)$ A graph representation of how instances exchange toots in the Fediverse. Nodes are instances, and links indicate that two instances are federated.}
\end{figure*}

\pb{What is Federation in Pleroma?} Once a user joins an instance, they can follow \one other users on the same instance; or \two remote users from other instances. The latter action creates a federation link between two instances (\ie federation). Thus, remote toots are retrieved by the local server (using ActivityPub) and presented to the local users on their timelines.

Figure~\ref{fig:federation} illustrates an example of how these toots are subscribed to and retrieved across the federated network. Consider three instances \texttt{x.pleroma}, \texttt{y.pleroma}, \texttt{z.mastodon} with users [Alice, Bob], Carol, and Dave, respectively. 
Let us assume \texttt{*.pleroma} hosts Pleroma and \texttt{*.mastodon} hosts a Mastodon microblogging service. ActivityPub allows Alice to follow Carol, who is a user on a different instance. While this is not possible with traditional platforms (\eg a user on Twitter cannot follow a user on Facebook), the Fediverse allows this flexibility. This is done by Alice performing a remote follow request to Carol which involves the following steps.
\circled{1} Alice makes a request to her local instance (\texttt{x.pleroma}) to follow Carol (on instance \texttt{y.pleroma}).
\circled{2} The request is forwarded to \texttt{y.pleroma}. 
\circled{3} The remote instance (\texttt{y.pleroma}) informs Carol that Alice is now following her. 
Thereafter, whenever \circled{4} Carol posts a toot on her instance (\texttt{y.pleroma}), \circled{5} it gets pushed to Alice's instance (\texttt{x.pleroma}).
Finally, \circled{6} when Alice logs in, the toot will appear on her timeline. 
Note, Alice can also view the video posted in \texttt{z.pleroma} as Bob (also from \texttt{x.pleroma}) remotely follows Dave (from  \texttt{z.mastodon}). 
This federated approach results in a complex network, where instances are linked as a result of the following relationships of their user base.

\pb{Challenges in Toxic Content Moderation.} 
Administration in the Fediverse is decentralised: each instance decides which toots are considered toxic \vs non-toxic.
Thus, prior centralized approaches to moderation, where a single administrative entity (\eg Twitter) has full control, no longer applies. This greatly complicates moderation as toots generated on one instance can easily spread to another instance, even if those two instances have wildly different viewpoints. 
For example, imagine \texttt{z.mastodon} allows the posting of pornographic content, yet \texttt{x.pleroma} does not. An explicit post by Dave --acceptable in his instance (\texttt{z.mastodon})-- may easily spread to \texttt{x.pleroma} where it is not acceptable.
Given the huge scale of the Fediverse and the voluntary nature of most administrators, manual moderation of large quantities of content is not feasible. 
This would makes it difficult for \texttt{x.pleroma} to moderate all incoming posts from \texttt{z.mastodon}.
Complicating this is the fact that instance administrators rarely wish to upload content to centralised moderation APIs, as this naturally undermines the nature of the DW (and incurs extra costs).
To give a preliminary sense of the scale of this federation network, Figure~\ref{fig:pleroma30connection} presents the network representation of the federation links formed between the instances in our dataset (later explained in \S\ref{sec:dataset}). 
Preventing the spread of toxic content with this dense and complex instance interconnectivity is challenging due to the ability for content to spread across the federated links.

\section{Dataset and Methodology}
\label{sec:dataset}

We start by describing our data collection and toxicity labeling methodology.
Although the Fediverse contains a large number of diverse services, we focus on Pleroma (microblogging platform). We do this because it is one of the largest by both content and user counts. 
Our data collection follows three key steps: 
\one~Discovering a set of Pleroma instances to measure, including its federation network;
\two~Gathering the full set of toots from each instance;
and
\three Labeling each toot as toxic or non-toxic. For the latter, we use \emph{toxicity} annotations from Jigsaw Perspective~\cite{perspective}, due to its widespread uptake and well-understood definition.

\pb{Discovering Instances.}
We first need to identify the domains of Pleroma instances deployed around the globe. To do this, we crawled a list of Pleroma instances from \url{the-federation.info/pleroma} on December 15, 2020. This yielded 729 unique instance domains.
We then expanded this list by recursively capturing the instances that these instances federate (\ie its list of remote instances it has previously connected to). This is done using each instance's Peers API\footnote{<instance.uri>/api/v1/instance/peers} between December 2020 and January 2021.
This API endpoint returns a complete list of instances that any instance has federated with during its lifespan. In total, we identify 1360 instances.

\pb{Collecting User Data \& Toots.}
Next, we collect all public toots from the identified instances. For this, we gathered all toots using their Public Timeline API.\footnote{<instance.uri>/api/v1/timelines/public} 
This API endpoint returns all public toots on the instance.
Each toot includes information about the author, text content, associated media, timestamp, number of likes, number of reblogs, and any self-tagged content warnings.
The latter is voluntarily added by the toot author to notify future viewers that the toot may contain (subjectively judged) sensitive material. In total, 5.4\% of toots in our dataset are self-tagged with warnings. 

To gather this data, we build and execute a multi-threaded crawler to  gather all prior toots made before January 22, 2021, from all responsive instances. This covered 713 (out of 1360) instances.
The primary reason for failures among the remaining instances was that many instances were not reachable (31.5\%), and some (12.3\%) had zero toots. The remaining instances did not make their toots publicly available and we made no
attempt to circumvent the publicly available data restrictions. Finally, we gather associated user profile information, namely the full list of each account's followers and followees.
Note, due to the distributed nature of the instances, we parallelised our crawler across several servers and implement a set of precautions, like rate limiting on API requests, to not burden the instances.

\pb{Toxicity Labels.}
This paper focuses on exploring the spread of toxicity on Pleroma. Hence, we label toots in our dataset using Google Jigsaw’s Perspective API~\cite{perspective}.
Our choice of Perspective API is motivated by similar and recent measurement studies of the other social platforms like 4chan~\cite{papasavva2020raiders} and Voat~\cite{papasavva2021qoincidence}. We apply the same classification model and toxicity definition for comparability. Perspective defines toxicity as ``a rude, disrespectful, or unreasonable comment that is likely to make people leave a discussion``. Internally, Perspective trains BERT-based models~\cite{devlin2018bert} on millions of comments from online forums such as Wikipedia and The New York Times where each comment is tagged by 3-10 crowdsourced annotators for toxicity. For a given toot, Perspective returns a score (between 0 and 1) for its toxicity.
This offers an estimate of the fraction of human moderators who would label the content as toxic.
Following~\cite{rottger2020hatecheck,papasavva2020raiders,papasavva2021qoincidence,reichert2020reading}, we consider a \emph{toot} to be toxic if its toxicity score is greater than 0.5 (and vice versa).
This represents a moderately toxic score, but we opt for this to capture a broad range of problematic content and behavior.
We label a \emph{user} as toxic if the average toxicity of their toots is greater than 0.5 (and vice versa). 
For completeness, we also repeat our later experiments with a stricter toxicity threshold (0.8) to find that it provides similar results, as summarised in the Appendix \ref{sec:additionalexperiments}. 

\begin{figure}
     \centering
     \subfloat[][]{
     \includegraphics[width=.5\linewidth]{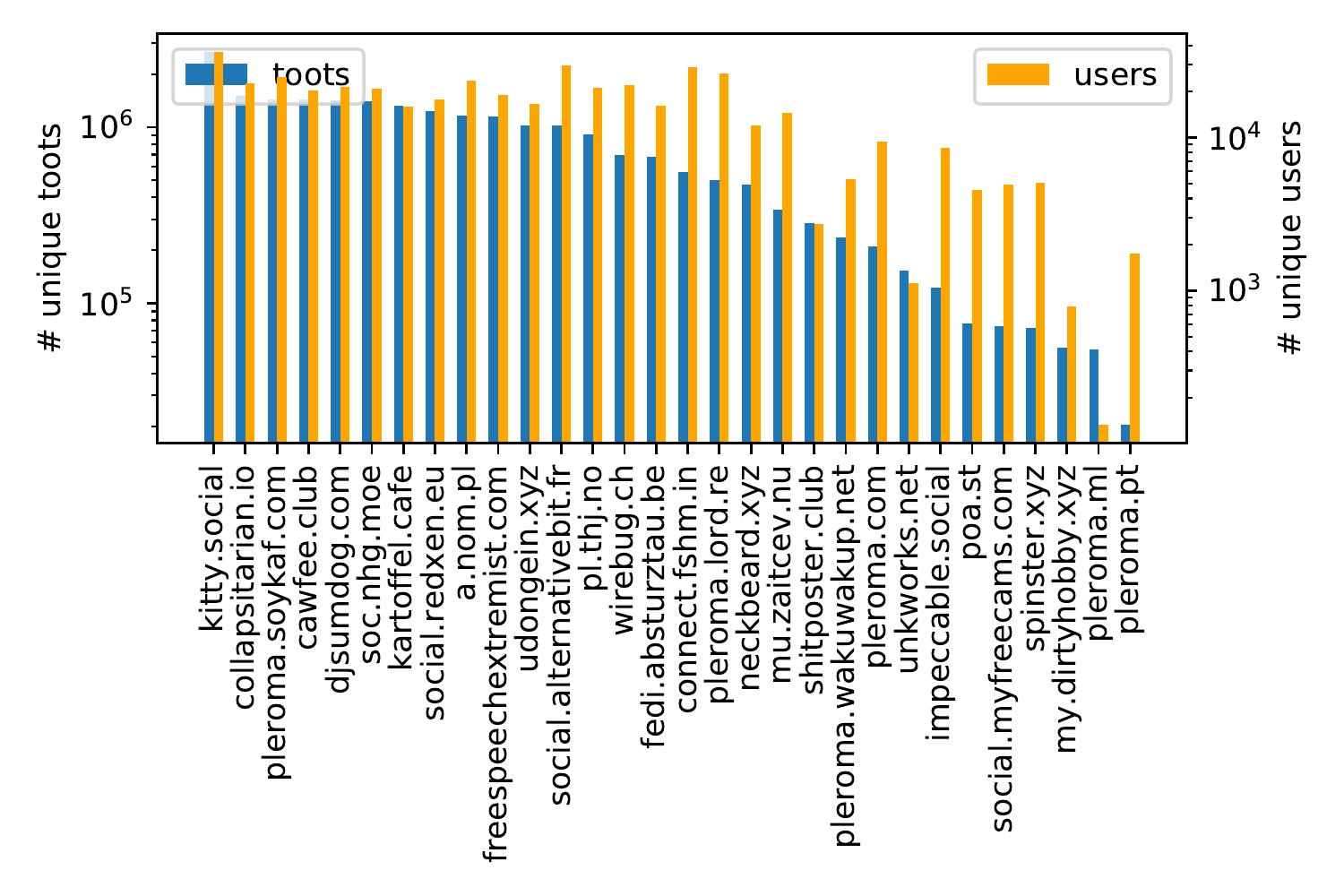} \label{fig:allusertootinfo}}
     \subfloat[][]{\includegraphics[width=.4\linewidth]{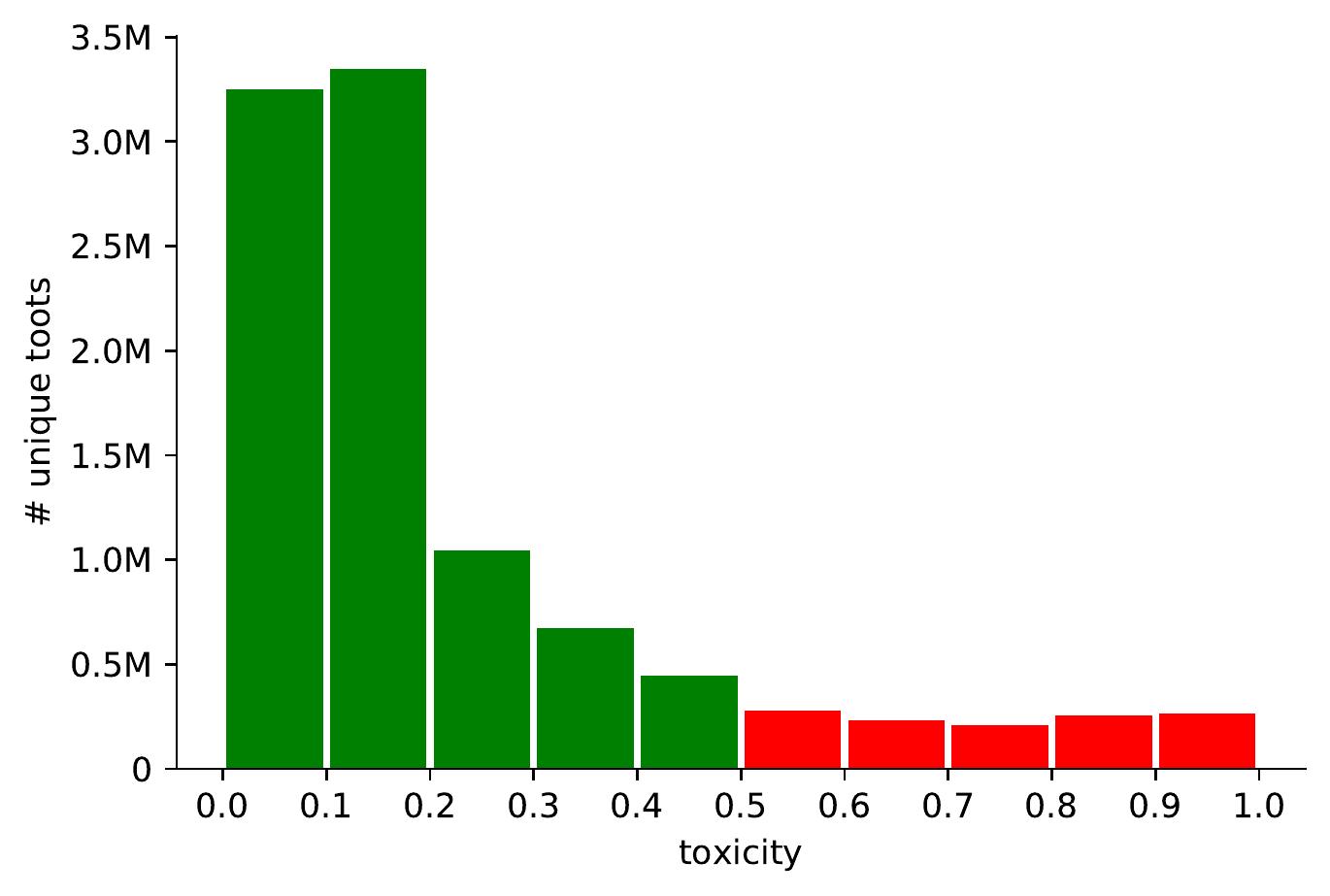}\label{fig:toxicityhistogram}}
     \caption{(a) Distribution of toots and users on Pleroma instances (note log scale on Y-axis). (b) Number of toots in different toxicity intervals.}
     \label{steady_state_1}
\end{figure}

Unfortunately, due to rate limitations, it is not possible to label \emph{all} toots in our dataset. Hence, we select the 30 largest instances based on the number of toots they contain (see Figure~\ref{fig:allusertootinfo}).
Overall, these 30 instances contain 9,927,712 unique toots (of which 1,394,512 were local toots) posted by 116,856 unique users (of which 8,367 were local users) between 1 January 2017 and 22 January 2021. 
This represents 55\% of the total toot count for that period. The federated graph showing the subscriptions of these 30 instances is shown in Figure~\ref{fig:pleroma30connection}. 
These 30 instances are connected to many further instances (on average 1,511). This implies that the annotated toots in our dataset include many toots that have been imported from  other instances.

In \S\ref{sec:classifiers} we use these Perspective annotations in lieu of human labels provided by administrators of each instance.
A limitation of this methodology is that each instance will obviously have annotations using a consistent and standardised definition (as dictated by Perspective). 
This is not necessarily representative of where individual instance administrators may moderate on differing criteria. 
To address this, we later introduce randomised noise into the annotations on a per-instance basis, to reflect discrepancies between administrators' annotation styles.

%
%

%
%
%
%
%

%

%

\section{Characterising Toxicity in Pleroma}
\label{sec:measurements}

Considering recent media reports of toxic activity in the DW~\cite{terror}, we start by presenting a general overview of the toxicity in Pleroma. We use this to quantify and motivate the importance of moderation.

\subsection{Overview of Toxicity}

The toxicity labels obtained from Perspective range between 0 and 1. %
Figure~\ref{fig:toxicityhistogram} presents the distribution of toxicity scores across all toots. In line with studies of other social platforms~\cite{papasavva2020raiders,papasavva2021qoincidence}, the majority of toots in our dataset are non-toxic.
On average, we find the amount of toxic content is significantly less than that observed on other online fringe communities such as 4chan (37\%)~\cite{papasavva2020raiders} and Voat (39.9\%)~\cite{papasavva2021qoincidence}. That said, we still find a large portion of toxic toots (12.15\%).

We conjecture that this may be driven by certain instances  spreading large volumes of toxic content. 
Thus, Figure~\ref{fig:percentagetoxictoots} presents the percentage of toots on each instance that score above 0.5 toxicity. 
We find that 80.7\% of toots on \url{my.dirtyhobby.xyz} are classified as toxic. This adult-themed instance is classified as toxic primarily due to the abundance of sex-related conversations. Note, this also highlights the diversity of toxic content types considered in centralised moderation APIs.
In contrast, we observe that the majority of other instances have between 7--23\% of toxic toots, confirming a wide range of toxicity levels.
We do discover a small number of exceptions though, with 3 instances generating under 3\% toxic toots: 
\url{pleroma.ml} (2.2\%), \url{pleroma.wakuwakup.net} (0.8\%), \url{unkworks.net} (0.3\%).
This confirms that the DW \emph{is} composed of a variety of instance types, with certain instances generating a significant share of toxicity.

\subsection{Spread of Toxicity}

\begin{figure}
     \centering
     \subfloat[][]{
     \includegraphics[width=.45\linewidth]{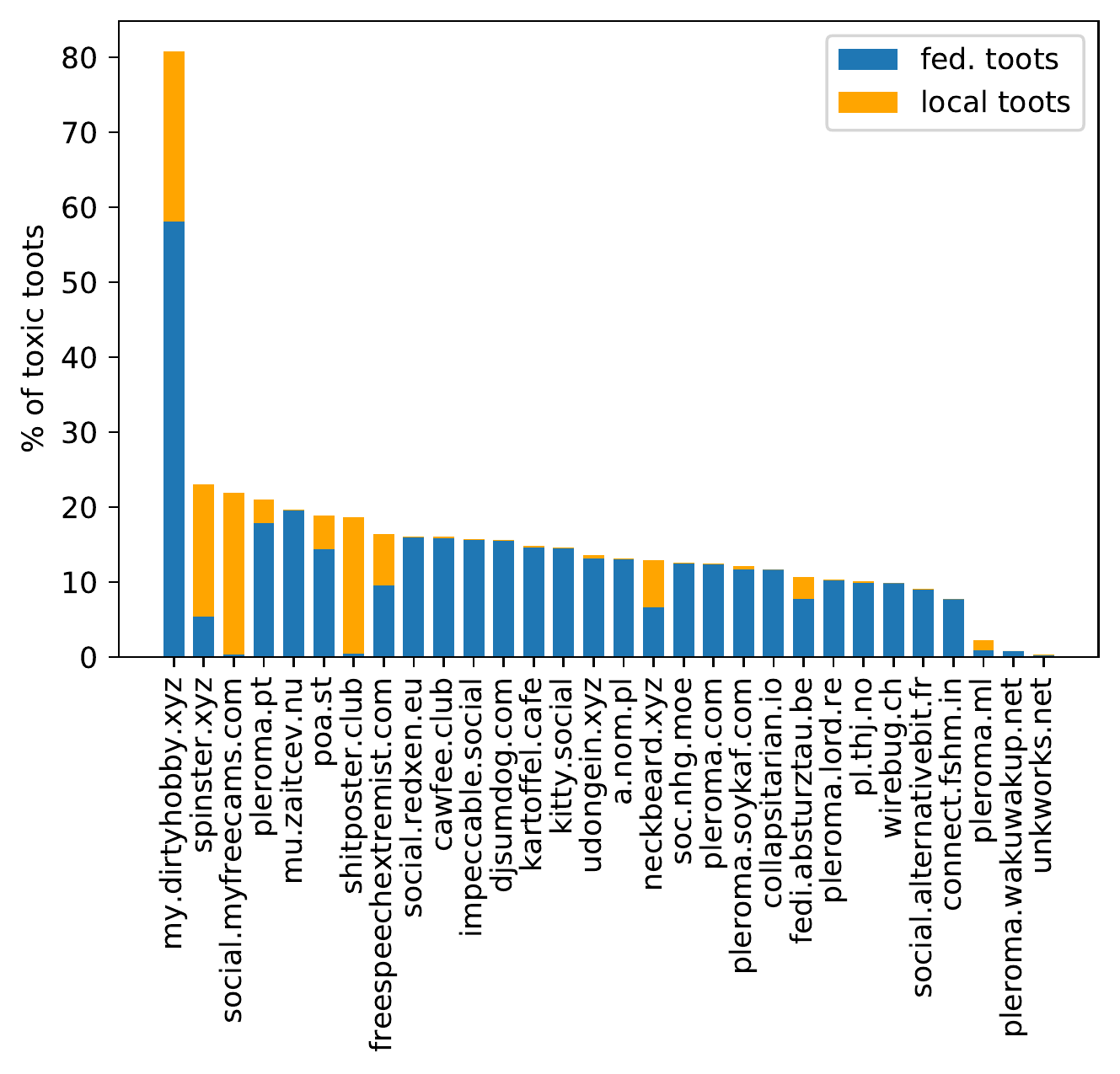} \label{fig:percentagetoxictoots}}
     \subfloat[][]{\includegraphics[width=.45\linewidth]{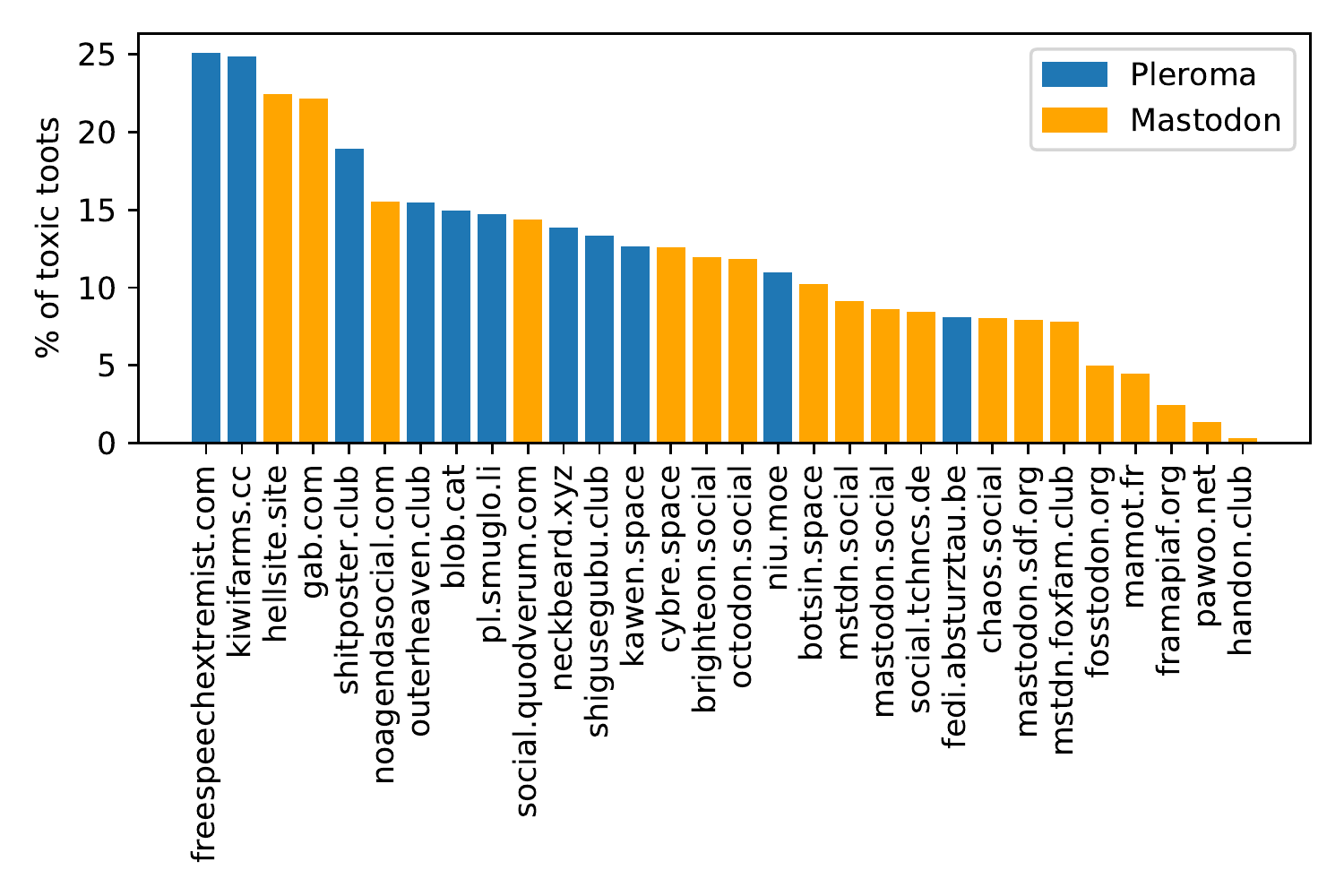}\label{fig:percentagetoxictootsonfedinstacnes}}
     \caption{Percentage of toxic toots (a) on Pleroma instances split by local and federated toots (b) from top-30 federated instances with the largest number of federated toots split by the origin of the federated toots.}
     \label{steady_state_2}
\end{figure}

A unique aspect of the DW is \emph{federation}, whereby content is imported from one instance to another. This poses a particular challenge for local administrators, who  have little control over the remote instances where content is created.

\pb{Local \vs Federated Toots.}
To test if this concern is legitimate, we compare the toxicity of local \vs federated toots (\ie toots that have been retrieved from a remote instance).
Figure~\ref{fig:percentagetoxictoots} plots the percentage of local \vs federated toxic toots on all 30 instances. In line with our intuition, we see that federated toots constitute the most significant chunk of toxic content on 26/30 of the instances. In contrast, only four instances have more toxic content generated locally (rather than imported via federation). Interestingly, this suggests that, on average, users tend to follow and import content that is more toxic than the locally generated one.

Next, recall that our dataset includes federated content that has been retrieved from instances outside of Pleroma (the W3C ActivityPub~\cite{activitypub} protocol allows Pleroma instances to interoperate with various other DW-enabled platforms such as Mastodon). 
In total, our 30 instances have federated with 5,516 other instances from the wider Fediverse.
This allows us to inspect the composition of toxic \vs non-toxic toots contributed by these other platforms.
To this end, we extract the set of instances (from the 5,516) with at least 10\% of their toots classified as toxic. This results in 1,489 non-Pleroma instances (27\%) being extracted. 
Figure~\ref{fig:percentagetoxictootsonfedinstacnes} plots the percentage of toxic toots from these federated instances with the largest number of federated toots. 
We observe a mix of both Pleroma and Mastodon (the two most popular DW microblogging services).
For instance, we see prominent controversial Pleroma instances like \url{kiwifarms.cc}, an instance that manages various forms of group trolling, harassment, and stalking.
We also observe several large Mastodon instances like \url{gab.com}, famed for hosting hate speech material~\cite{zannettou2018gab, gabblog}.
Interestingly, not all federated instances are equally harmful in absolute terms. We observe a significant number of instances that contribute a relatively small number of federated toots ($<100$ per instance), even though a significant fraction of them are toxic. %
This confirms that federation \emph{is} a significant challenge for per-instance moderation: regardless of how well an administrator moderates their own instance, it is possible for millions of remote (toxic) toots to be retrieved by users from other instances that may have very different policies. The openness of DW federation exacerbates this further, by allowing different platforms (\eg Pleroma, Mastodon, PeerTube) to interoperate. Thus, any solutions cannot rely on remote instances adhering to identical practices, as they may be running distinct software stacks.

\begin{figure}[t]
\centering
\includegraphics[width=\textwidth]{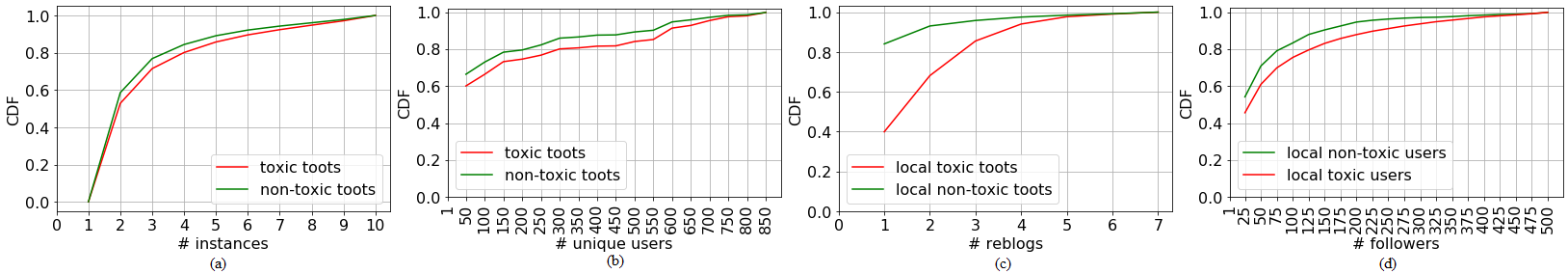} 
\caption{CDF of (a) the number of instances federated toots reach split by toot toxicity, (b) number of users federated toots reach split by toot toxicity, (c) number of reblogs of toxic and non-toxic toots, (d) number of followers of toxic and non-toxic users.}
\label{fig:allcdfs}
\end{figure}

\pb{Reach of Toxic Toots.}
After showing how toxic content flows across instances, we now study how many instances and users this content spreads to.
Figure~\ref{fig:allcdfs}a shows the number of instances that toots are replicated onto (via federation). 
The distributions are extremely similar, with a Kolmogorov-Smirnov p-value of 0.85.
However, we do see that toxic content has a marginally higher probability of reaching a larger number of instances (than non-toxic content).
In total, 52\% of toxic toots are federated on at least one other instance, and, on average, toxic toots are replicated onto three instances \vs  two for non-toxic toots.

This, however, may be misleading as different instances have different user populations sizes. For example, a toot being replicated onto 10 instances, each with one user, has less reach than a toot being replicated onto a single instance with a million users.
To examine the audience reach in terms of the number of users the content hits, Figure~\ref{fig:allcdfs}b plots the number of unique users each toot reaches. We calculate this based on the number of registered users on each instance. Here, we observe clearer trends, with a Kolmogorov-Smirnov p-value of 0.24.
This confirms that toxic toots reach noticeably larger audiences. 87\% of non-toxic toots reach more than 1 user, compared to 92.3\% for toxic toots. On average, toxic toots appear on the timelines of 1.4x more users.

\subsection{Characterising Local Toots \& Users}

We next seek to explore the characteristics of local toots and users that are toxic \vs non-toxic. For this analysis, we therefore exclude all federated toots. In total, this leaves 1,394,512 unique local toots (of which 17.6\% are toxic). These are posted by 8,367 unique local users (of which 12.7\% are considered toxic). By focusing solely on local toots, we can better understand the characteristics of each individual instance's user base.

\pb{Reblogs.}
We first inspect the \emph{reblog} rate of toxic toots (a reblog is equivalent to a retweet on Twitter).
We conjecture that toxic toots are likely to get more reblogs than non-toxic ones. To explore this, we analyze the number of reblogs that toxic \vs non-toxic toots receive.
Figure~\ref{fig:allcdfs}c presents the CDF of the number of reblogs observed. 
Indeed, we see that toxic toots gain substantially more reblogs. 60\% of toxic toots get more than one reblog, compared to only 16\% of non-toxic toots. 
This trend indicates that interest and uptake in toxic material are consistently greater. 
On average, toxic toots get 140\% more reblogs than their non-toxic counterparts. This confirms the virality of toxic content and helps explain the greater spread of toxic toots.

\begin{table}[t]
\small
\centering
\resizebox{\columnwidth}{!}{%
\begin{tabular}{lcccc}
\toprule
\textbf{Topic} &
  \textbf{Representative words} &
  \textbf{\begin{tabular}[c]{@{}c@{}}Distribution\\ in all toots\\ (\%)\end{tabular}} &
  \textbf{\begin{tabular}[c]{@{}c@{}}Distribution\\ in toxic toots\\ (\%)\end{tabular}} &
  \textbf{\begin{tabular}[c]{@{}c@{}}Toxic toots\\ per topic\\ (\%)\end{tabular}} \\ \midrule
General conversation            & time, like, know, date, message                    & 29.5 & 28.5 & 18.8 \\ %
Profinity                       & fuck, wtf, ass, shit, holy                         & 8.8  & 22.0 & 46.2 \\ %
Dark web                        & dark, net, deep, box, web                          & 7.4  & 4.4  & 11.2 \\ %
Sex talks and online cams       & online, love, video, cum, guys                     & 7.0  & 7.9  & 21.7 \\ %
Computers, e-games and tech.    & game, linux, work, windows, play                   & 7.0  & 4.9  & 13.3 \\ %
Greetings and compliments       & hope, good, day, morning, nice                     & 6.9  & 4.4  & 12.1 \\ %
Fediverse                       & post, server, mastodon, pleroma, instance          & 6.7  & 4.0  & 11.4 \\ %
NSFW content                    & jpg, nsfw, source, tits, porn                      & 5.4  & 0.9  & 3.3  \\ %
COVID and economy               & coronavirus, million, economics, pandemic, markets & 4.9  & 1.5  & 6.0  \\ %
Loli content                    & png, image, loli, screenshot, husky                & 4.7  & 1.6  & 6.5  \\ %
Politics                        & trump, election, vote, biden, america              & 4.4  & 7.2  & 31.4 \\ %
Humam rights (esp. gender)      & women, men, gender, white, rights                  & 3.7  & 9.1  & 46.3 \\ %
Free speech and societal issues & governance, people, speech, crowd, free            & 3.6  & 3.6  & 18.7 \\ \bottomrule
\end{tabular}
}
\caption{Topics in local toots as determined by CombinedTM and interpreted using pyLDAvis with their overall distribution, distribution in toxic toots, and percentage toxic toots per topic.}
\label{tab:topicanalysis}
\end{table}

\pb{Followers.}
Pleroma allows its users to follow other users across the Fediverse. 
In addition to toots, we also have information about each user's follower list. 
We next inspect whether there is a relationship between the number of followers a user has and the toxicity the account produces.
Figure~\ref{fig:allcdfs}d plots the CDF of the number of followers of toxic and non-toxic users. Recall, we define a user as toxic their average toxicity score exceeds 0.5.

We observe that toxic users tend to have more followers, with a Kolmogorov-Smirnov p-value of 0.96. On average, toxic users have 142 followers compared to just 70 for non-toxic ones.
One possible explanation is that this may be driven by differences in the frequency of toots (as the users who toot frequently are more likely to be viewed). 
To test this, we compute the toot:user ratio for toxic \vs non-toxic users.
This actually shows a contrary trend.
On average, non-toxic users have 188 toots per user compared to just 28 for toxic users. This suggests that our population of non-toxic users are actually more active. 
Therefore, we conjecture that these patterns may be driven by the higher reblog rates for toxic toots (see Figure~\ref{fig:allcdfs} (c)), thereby increasing the exposure of such accounts.

\pb{Distribution of Topics.}
Until now, we have analysed toots regardless of their instance. However, users on different instances may hold discussions on very diverse topics with varying levels of toxicity. 
This may play a large role in defining and identifying toxicity. 
Thus, we analyze the topics found in local toots using topic modeling to get a more granular view~\cite{ramage2009topic}.

For this, we use CombinedTM from Contextualized Topic Models (CTM)~\cite{bianchi2020pre}, which combines contextual embeddings with the bag of words to make more coherent topics. The number of topics was chosen using a grid search over model coherence~\cite{roder2015exploring} and the model with the highest coherence was selected. In addition, we use pyLDAvis~\cite{sievert2014ldavis} to interpret topics and identify topic overlaps and similarity. Our final topic model results in 13 topics. We assign each toot with the most probable topic as predicted by the topic model. Table~\ref{tab:topicanalysis} lists the final topics, their overall distribution, distribution in toxic toots, and percentage of toxic toots per topic. 

We observe that instances participate in diverse discourse ranging from sex to politics. Interestingly, the make-up of these conversations (in terms of toxicity) differs across each topic.
Unsurprisingly, general everyday conversations form the largest portion of both overall (29.5\%) and toxic (28.5\%) toots. We also see several contrasting topics, where they contribute a larger share of toxic toots as compared to the overall distribution. For instance, human rights contribute just 3.7\% of toots yet constitutes one of the most significant portions of toxic content (9.1\%). In our dataset, this topic primarily covers gender issues, with strongly worded dialogue throughout.
Similar comments can be made for Profanity (8.8\% overall \vs 22.0\% of toxic toots) and Politics (4.4\% overall \vs 7.2\% toxic). The former is not surprising as, by definition, profane toots are classified mainly as toxic. However, it is perhaps more worrying to see the significant density of toxic behaviour when discussing politics. As our dataset covers the 2019 United States Presidential Elections, we find extensive discussion about people such as Trump and Biden. 
This highly polarising topic triggered substantial confrontational and abusive language in our dataset. We again emphasise that our annotations are based on Perspective, providing a unified definition of toxic across the instances. In practice, we highlight that individual instance administrators may have differing views on what they personally consider toxic.

\begin{figure}[t]
\centering
\includegraphics[width=\textwidth]{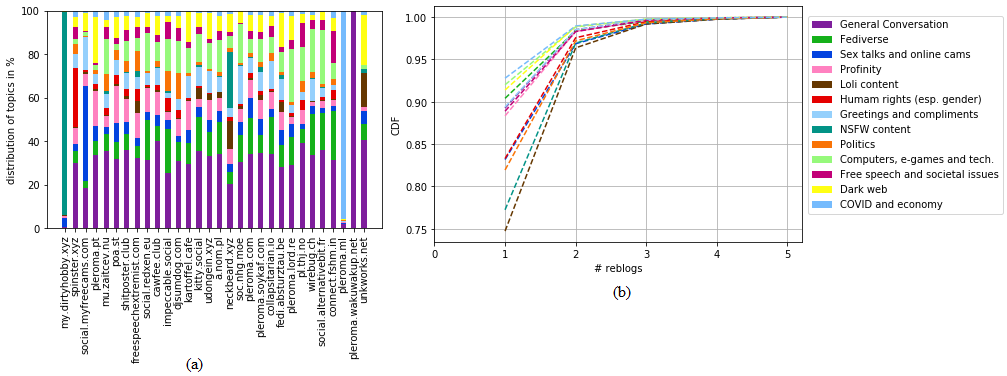} 
\caption{(a) Distribution of topics on each instance, and (b) CDFs of the number of reblogs of toots belonging to each topic.}
\label{fig:alltopics}
\end{figure}

\pb{Topics \vs Instances.}
As topics may not be evenly distributed across all instances, we also analyse the distribution of topics on individual instances.
We conjecture that some of the above trends may be driven by a subset of controversial instances. For example, a larger number of toots on a given topic could simply be generated by a single highly active instance.
Figure~\ref{fig:alltopics}a presents the percentage of toots belonging to each topic on a per-instance basis. We see that the topics discussed across instances \emph{do} vary. Other than the general day-to-day discussions that occur almost everywhere, most instances prefer some topics more than others. For example, Not Safe for Work (NSFW) content is popular on \url{my.dirtyhobby.xyz} and \url{neckbeard.xyz}, whereas human rights (particularly gender issues) is discussed heavily on \url{spinster.xyz}.
This flags up noticeable challenges for automating the detection of toxic content on the instances.
This is because many toxic content models fail to generalize well beyond their target environment~\cite{vidgen2019challenges,vidgen2020directions}. Thus, applying models trained on human rights discussions to gender issues may not transfer well.

To further probe into the popularity of these topics, we also analyze the reblogs of toots belonging to each topic. Figure~\ref{fig:alltopics}b shows the CDFs of the number of reblogs of toots split by topic. We observe that Loli content (sexualised anime material) receives more reblogs than any other topic, followed by NSFW toots. Again, this further confirms the virality of sensitive content on Pleroma instances.

\section{Exploring Automated Moderation}
\label{sec:classifiers}

The previous section has confirmed the presence of large volumes of toxic material and its propensity to spread further and faster than non-toxic content. This is a challenge for administrators who must moderate not only their own instance's toots but also any federated material imported by their userbase. As the use of centralised moderation API (\eg Perspective) undermines the decentralisation goals of the DW (as well as introducing cost, privacy and overhead issues), we argue that operating locally trained classification models is the only way forward. 
Thus, we next explore the potential of deploying automated content moderation on instances. 

\subsection{Experimental Methodology}

As instances do not have in-built content classification tools, we propose and evaluate several potential architectures. 
We first take the toxicity annotations listed in \S\ref{sec:dataset} to train local models for each instance. These annotations include both the Perspective labels (recall that 12.15\% of all toots are toxic) \emph{and} the self-tagged content warnings for each toot (recall that 5.4\% of all toots contain these). Note, we use the Perspective labels in lieu of human decisions made by an instance's administrator.

We next train models for each instance using their respective toot datasets.
Recall that the data of each instance includes both the toots from the local users as well as the federated toots 
from remote users followed by the local ones (see \S\ref{sec:background}).
Numerous text classifiers could be used for this purpose. 
Due to the resource constraints of Pleroma instances (many run on Raspberry Pis), we choose the methodology used by two heavily cited seminal works~
\cite{davidson2017automated,wulczyn2017ex}.
Namely, we rely on a Logistic Regression (LR) classifier with bag-of-words features~\cite{berger1996maximum}, implemented using Scikit-learn~\cite{pedregosa2011scikit}. 
We also experimented with an SVM classifier and obtained equivalent results (as measured using Student's t-test with p $>$ 0.05). We report these additional experiments in the Appendix.\looseness=-1 Note, we acknowledge that this simple model although resource-light has limitations \eg it discards word order and context and in turn could not differentiate between the same words differently arranged ("you are stupid" vs "are you stupid").

\subsection{Performance Across Instances}
\label{sec:performanceacrossinstances}

First, we compare the performance of the models trained on the two different label schemes (content warnings \vs Perspective). We train per-instance models based on all the toots available to it, with an 80:20 split stratified by respective labels between training and testing data. Our goal is to estimate the feasibility of using content warnings to help automate the annotation process.

\begin{figure}
     \centering
     \subfloat[][]{
     \includegraphics[width=.45\linewidth]{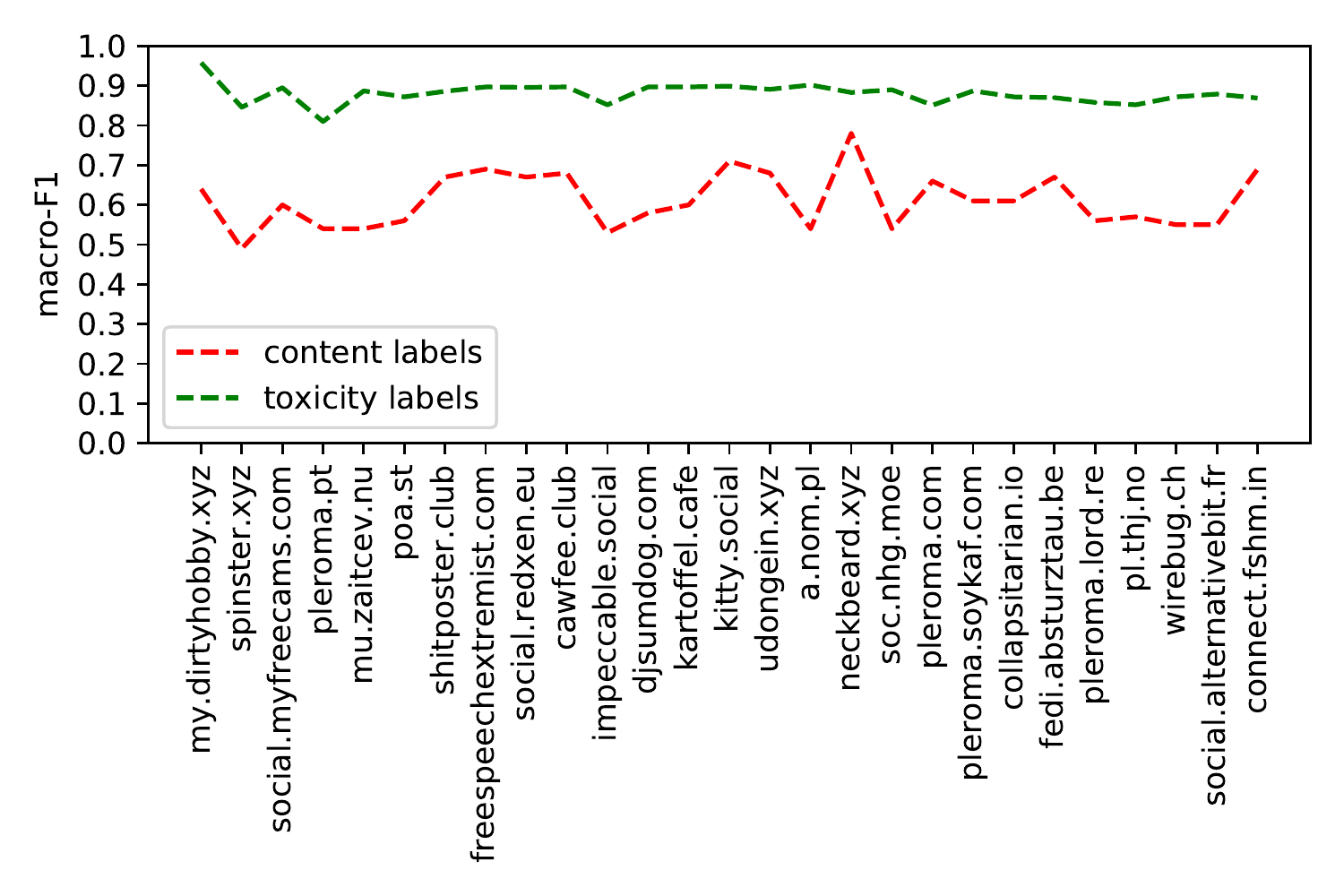} \label{fig:contenttoxicitylabels}}
     \subfloat[][]{\includegraphics[width=.45\linewidth]{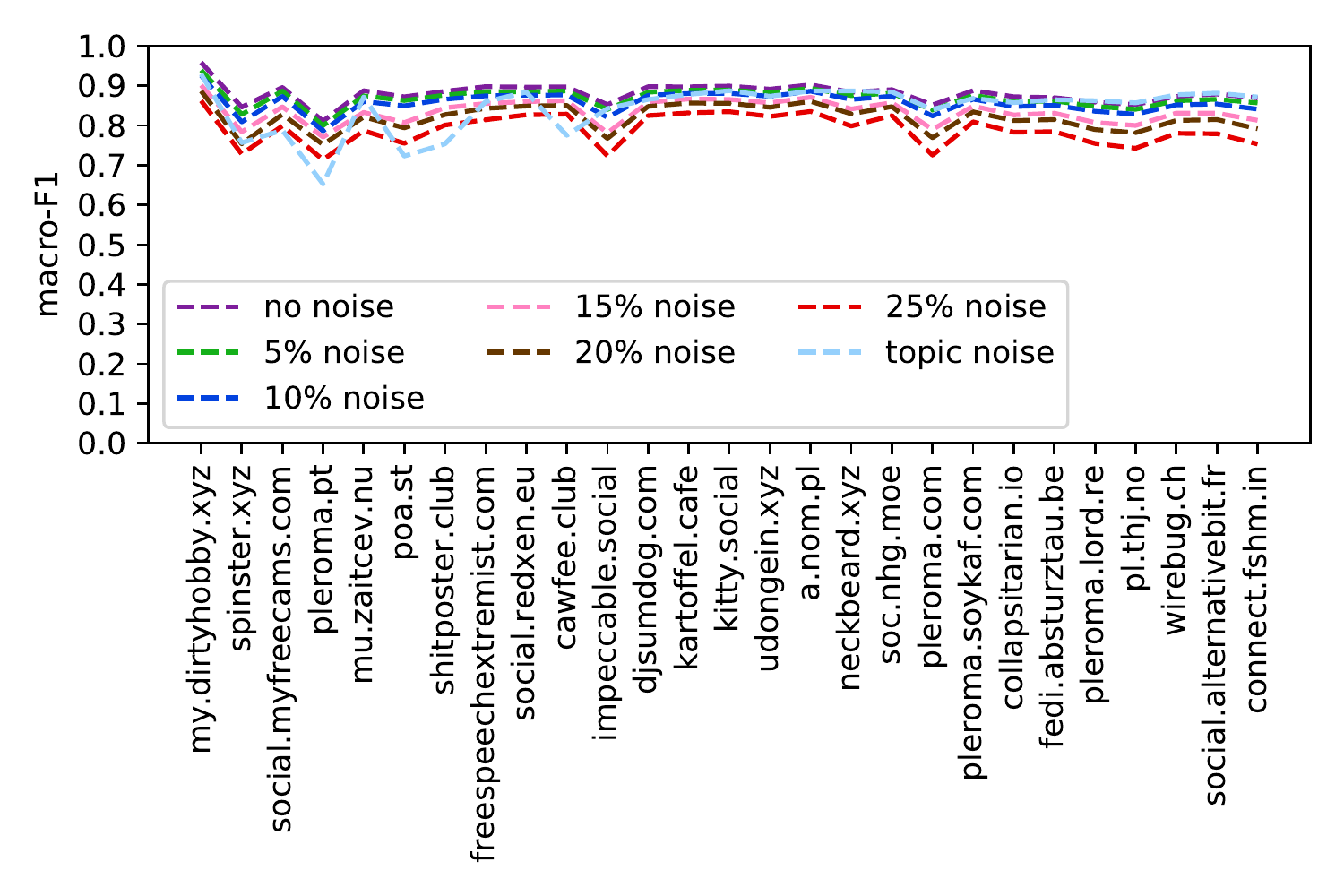}\label{fig:toxicitylabelsnoisy}}
     \caption{Macro-F1 scores of classifiers trained on (a) content labels vs toxicity labels for each instance, (b) toxicity labels with varying percentages of noisy labels for each instance.}
     \label{steady_state_3}
\end{figure}

\pb{Content Warning Labels.}
First, we use the self-tagged content warnings as annotations.
Figure~\ref{fig:contenttoxicitylabels} shows the macro-F1 scores of the classifiers for each instance. Overall, the results are not promising. The majority of the instances have a macro-F1 score less than or equal to 0.6, with only two instances exceeding 0.7.

We attribute this inefficiency to label noise introduced by inter-observer variability, \ie different users perceive sensitivity differently, which results in uncertainty of labels.
Also, we observe that these self-tagged content labels often do not necessarily correspond to typical interpretations of toxicity. 
For instance, we observe users adding  content warnings to their toots that reference news articles pertaining to war. 
To evidence this better, we calculate the inter-agreement between the content warnings and the toxicity labels from Perspective: the Cohen's Kappa is just 0.01.
In other words, there is only a small (random) chance that agreement exists between the two labels.

\pb{Perspective Toxicity Labels.}
The above indicates that using content warning labels is \emph{not} suitable for training local models.
Hence, we next repeat the previous training process for each instance using the Perspective labels instead. Figure~\ref{fig:contenttoxicitylabels} presents the macro-F1 scores of classifiers for each instance. The classifiers trained on Perspective labels show a significant improvement over those trained using content warnings. All the instances have a macro-F1 score greater than 0.8, with an average (calculated over all instances) of 0.84. This is 40\% more than the average of classifiers trained on content warnings.
This is reasonable as the Perspective labels exhibit far greater consistency than the self-tagged warnings.
We therefore confirm that it \emph{is} viable to train local models using data labeled with a consistent labeling scheme that can support and semi-automated content moderation for the administrators.

\pb{Variations in Toxicity Labels.}
One limitation of the above methodology is that the use of Perspective implies that annotations across all instances are identical. For example, if two instances have the same text in a toot, those two toots would be allocated identical Perspective labels. 
This does not necessarily reflect reality, as individual administrators may have different perceptions of what is ``toxic''. 
Administrators may also make mistakes or simply not exhibit consistent views across time. 

To assess the effect of annotation inconsistency, we emulate mistakes that might take place in the annotation process.
For each instance, we generate five new training sets with different noise levels (5\% to 25\%) by randomly flipping the labels of $x$\% of the toots (where $x=[5,25]$). We then repeat the above training process with these noisy labels.

Figure~\ref{fig:toxicitylabelsnoisy} presents the macro-F1 scores of the classifiers trained with varying degrees of noise in the labels.
We observe an expected gradual decrease in performance as the noise in the labels increases.
However, even with 25\% of noisy labels, the average (calculated over all instances) macro-F1 score decreases only by 11.9\%. This gives us confidence that it is feasible to build these local models, even in the presence of mistakes and inconsistency.

Whereas the above emulates mistakes, it does not reflect topical differences between annotation policies per-instance. Specifically, we expect that instances may have more systematic differences based on the theme of their instances. For example, an instance dedicated to sharing adult content is unlikely to tag sexual-related material as toxic. 
To assess the impact of this,  we generate a new topic-based training set for each instance.
For each instance, we select the topic that is the most popular among its users (excluding ``General Conversation'', see Figure~\ref{fig:alltopics}a).
We then whitelist all toots in that topic, and label them as non-toxic. For example, we select NSFW content for \url{my.dirtyhobby.xyz} and Human rights for \url{spinster.xyz}. 

Figure~\ref{fig:toxicitylabelsnoisy} presents the macro-F1 scores of the classifiers trained using this topic-based set. We see that these models perform far better than the prior experiments that introduced random noise. 
For these topic-based noisy models, the average performance decreases only by 4.7\%. 
This is largely because the local training sets retain consistency on what is and what is not toxic. 

\begin{figure}[h]
\centering
\includegraphics[width=0.7\columnwidth]{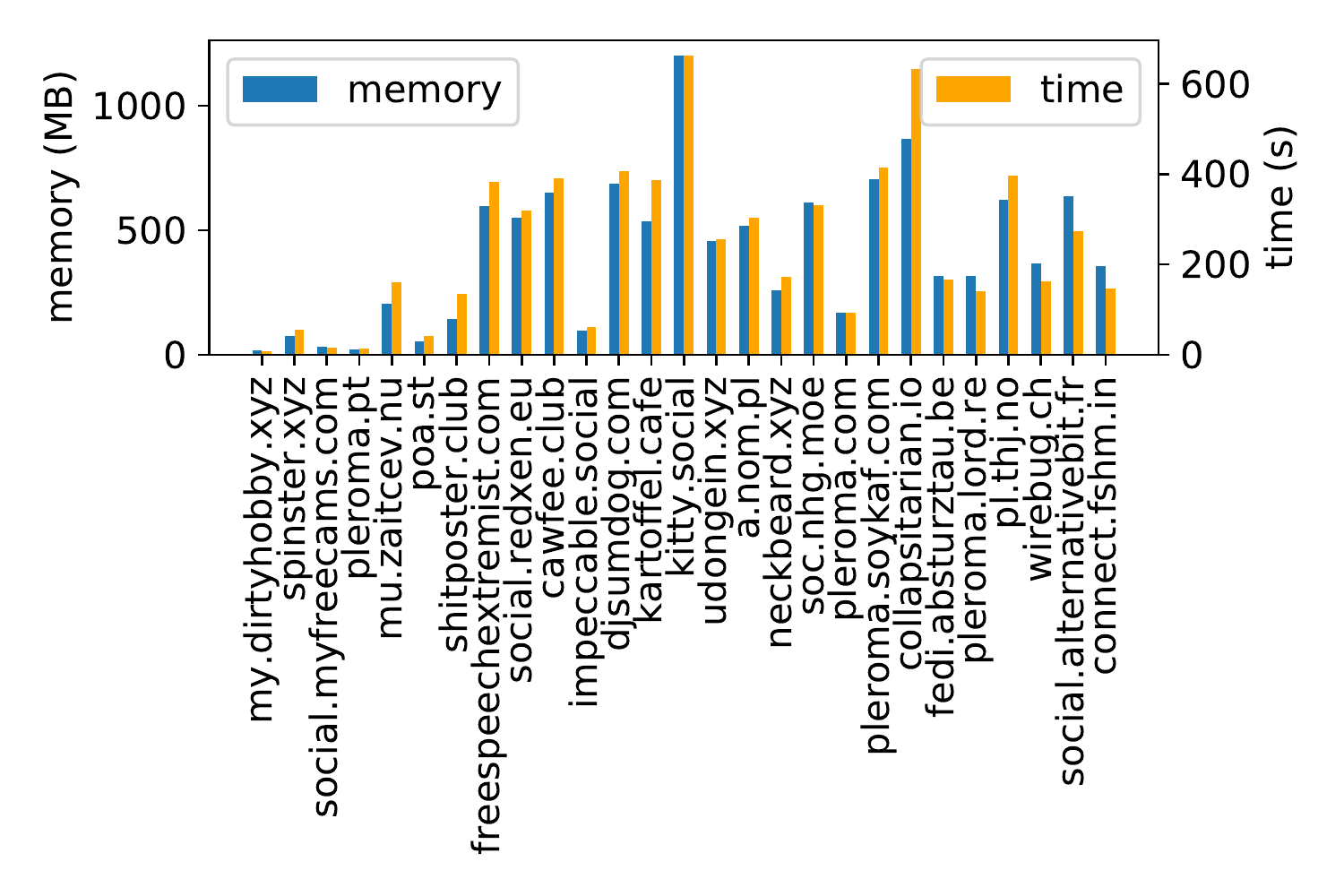} 
\caption{Memory \&  Time consumption of training local models on Pleroma instances.}
\label{fig:memoryandtime}
\end{figure}

\pb{Computational Time and Cost.} 
Due to the lightweight nature of each instance, for completeness, we finally plot the computational costs for training the models. 
We train each model on a machine with 128 GB of RAM with 16 cores (note, we did not use any GPUs). Figure~\ref{fig:memoryandtime}  shows the memory footprint of training each model, as well as the time taken. We see a wide variety of values, largely incumbent on the number of training toots within the instance. For example, \url{kitty.social} takes 662.3 seconds (containing 196,6617 toots), whereas \url{my.dirtyhobby.xyz} takes just 8.0 seconds (containing 29,987 toots). 
On average, training takes 241 seconds per model.
Importantly, we see that, on average, training only requires 409MB of RAM, suggesting this is well within the capacity of even lightweight hosted instances (\eg the Raspberry Pi 4 has 8 GB RAM).

%
%
%
%
%

%

%

%
\subsection{Annotation Feasibility}
\label{sec:annotation_feasibility}

The prior experiments have a key assumption.
The 80:20 split assumes that an administrator has time to annotate 80\% of toots on their instance. In practice, this is probably infeasible due to the voluntary nature of most Pleroma administrators.
Moreover, a newly created instance may only have annotated a tiny number of toots in its early days.
Thus, we next evaluate how many toots each admin must label to attain reasonably good performance.

For this, we generate new training sets of different sizes. 
For each instance, we extract the first $n$ toots (where $n=[500,10000]$ in intervals of 500) in the timeline, and  train 20 models (1x per set of size $n$). 
We also generate a second set, where we extract $n$ random toots from across the entirety of the timeline of each instance.
These two sets represent the cases where \one~Administrators dedicate their time to annotating all toots for the initial period of operation; and
\two~Administrators allocate occasional time to annotate a small subset of toots across the entire duration of the instance's lifetime.
In both cases, we repeat our prior training on each of these datasets and compute the macro-F1 on the test set of each instance.

\begin{figure*}
     \raggedright {
     \includegraphics[width=.75\linewidth]{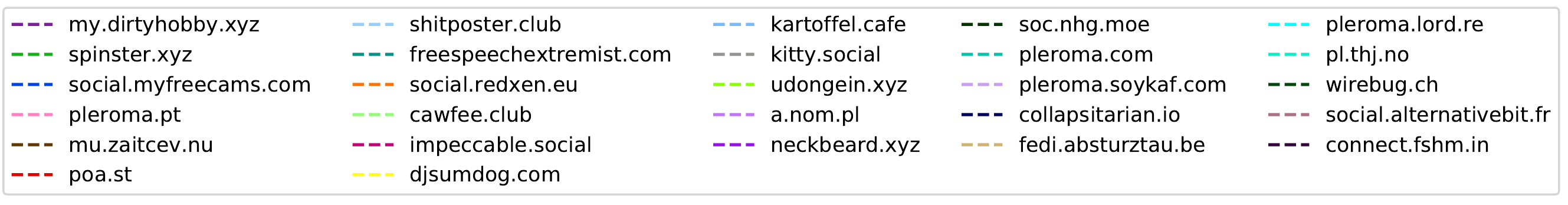}
     }
     \subfloat[][]{
     \includegraphics[width=.32\linewidth]{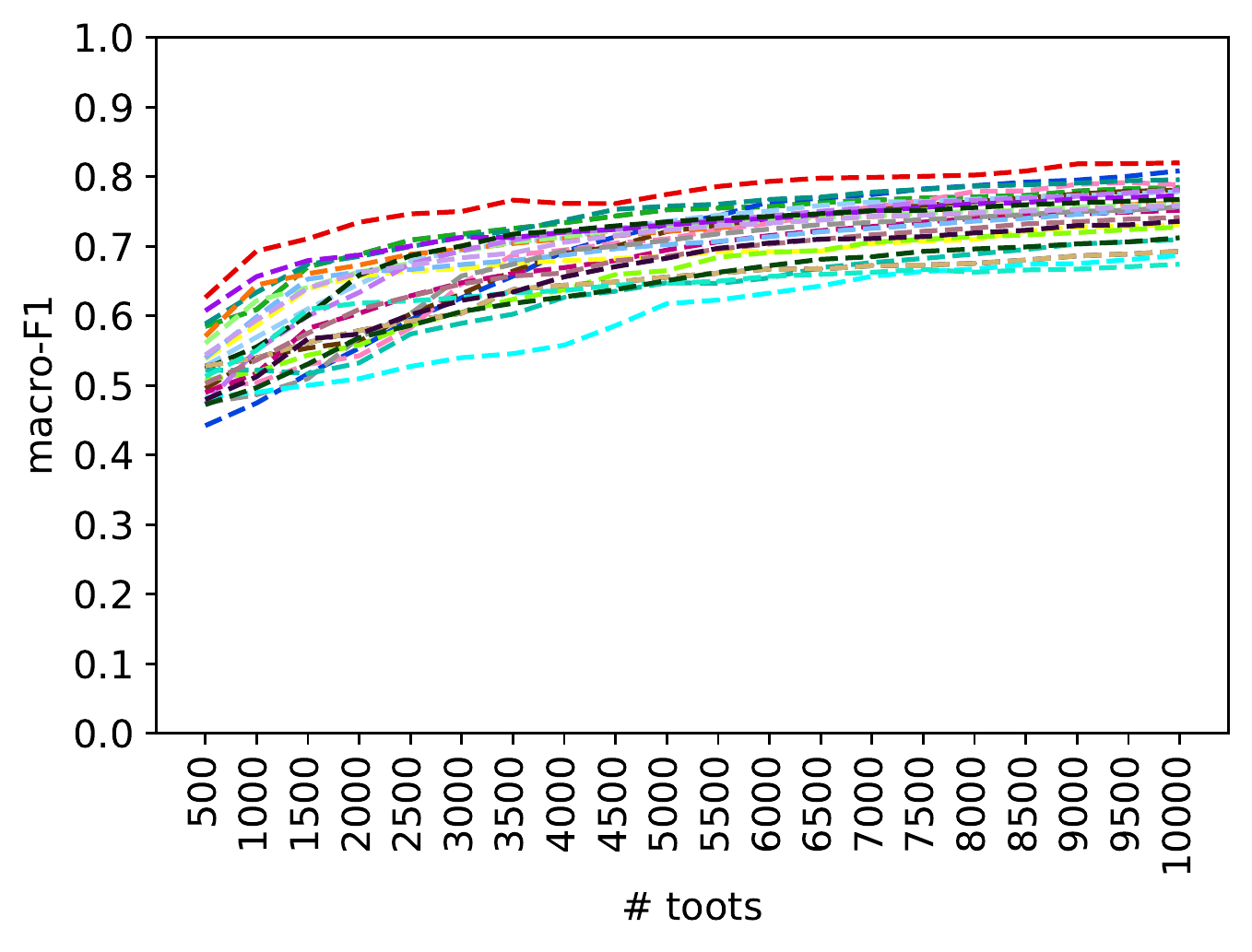} \label{fig:classifierN}}
     \subfloat[][]{\includegraphics[width=.32\linewidth]{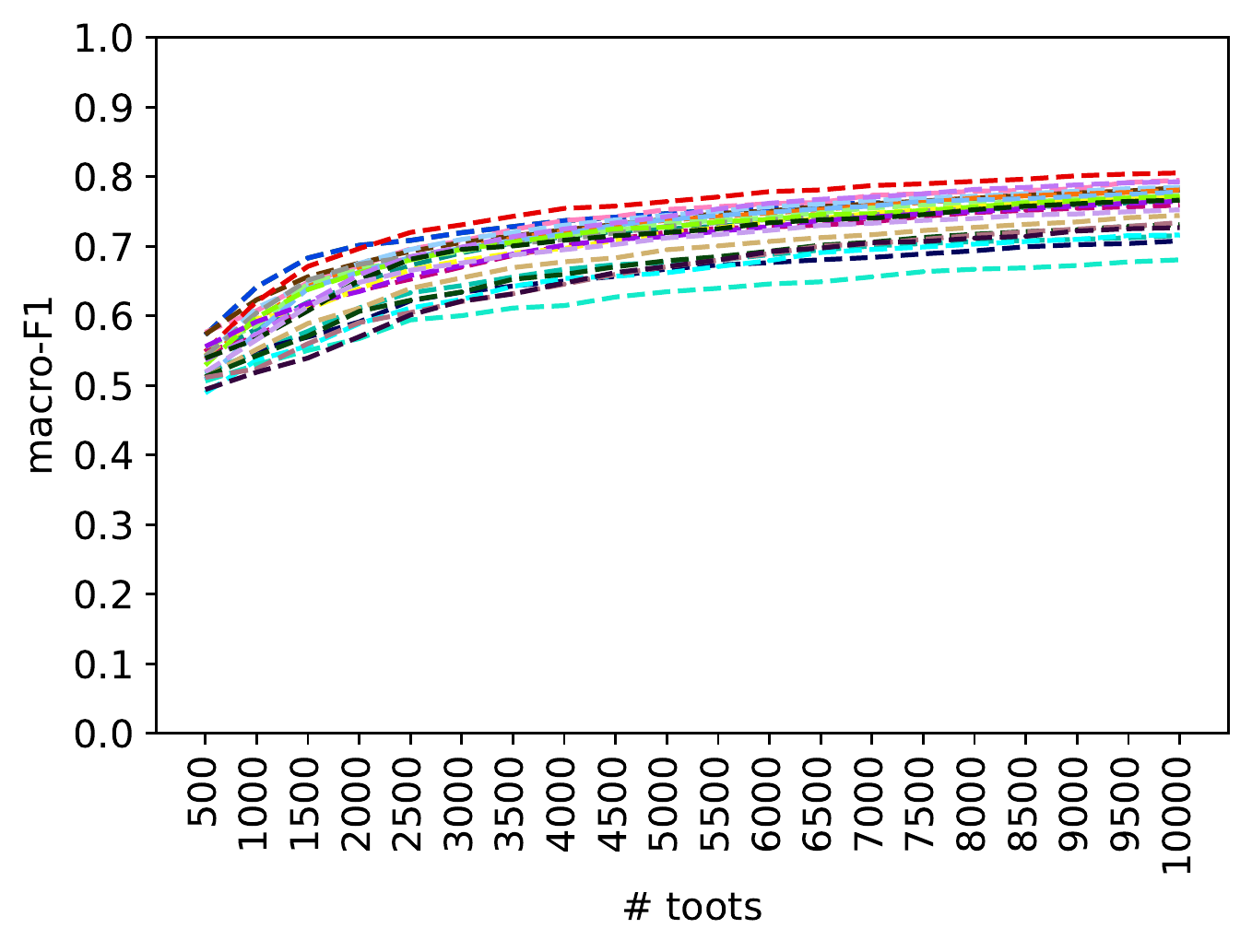}\label{fig:classifierNrandom}}
    \subfloat[][]{\includegraphics[width=.32\linewidth]{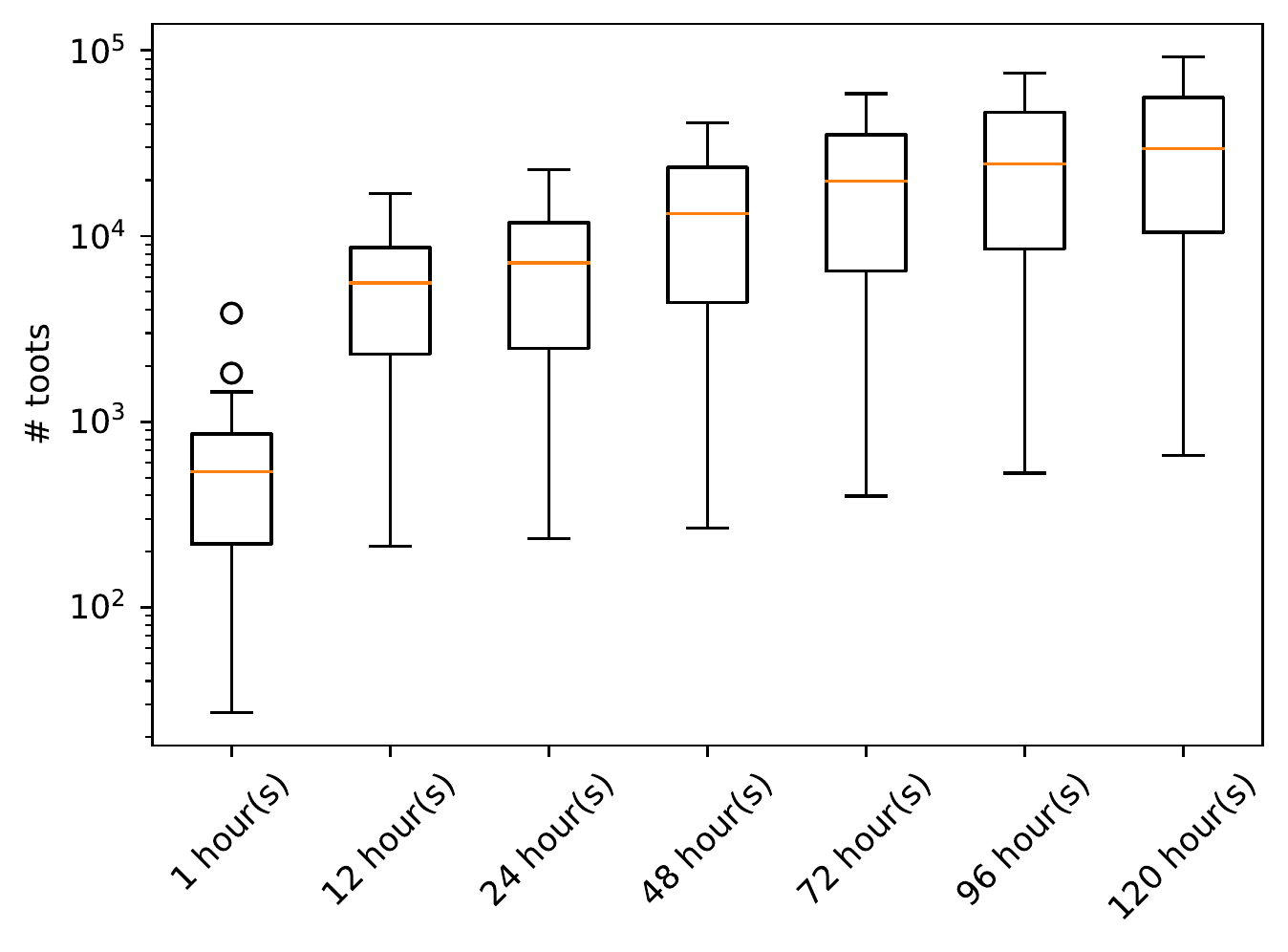}\label{fig:classifiertemporaltoots}}
     \caption{Macro-F1 scores of classifiers (a) trained on first $n$ toots, (b) trained on random $n$ toots. (c) Distribution of toots accumulated on all instances at different time steps (note log scale on Y-axis). }
     \label{steady_state_7}
\end{figure*}

Figure~\ref{fig:classifierN} shows the results generated from the first $n$ toots, whereas Figure~\ref{fig:classifierNrandom} shows the results for $n$ toots randomly selected from across the entire timeline. In both cases, the X-axis depicts the size of the training set.
We see similar results for all configurations. We observe an expected steady improvement in performance, with larger training sets (plateauing after around 6K posts). 
Although the trends are roughly similar for all instances, we do observe a range of performance and some outliers. For example, the models trained on \url{poa.st} achieve a relatively high macro-F1, starting with 0.62 on the first 500 toots and reaching a maximum of 0.81 on the first 10,000 toots.
On average, instances attain a macro-F1 of 0.52 on the first 500 toots and 0.75 on the first 10,000. The averages remain roughly the same when the classifiers are trained on the same number of toots selected randomly.

Unfortunately, these results show that instances, on average, need more than 10K labeled posts to get a reasonably good classification performance (0.80 macro-F1). %
This means it will be difficult for instances to train and deploy their own local moderation models for two key reasons. 
First, different instances accumulate toots at different rates.
To quantify this, Figure~\ref{fig:classifiertemporaltoots} presents a box plot showing the number of toots accumulated across different time periods.
This is calculated by taking the average number of toots generated over these time periods on a per-day basis for each instance.
We see a high degree of divergence (note the Y-axis log scale).
For example, obtaining a training set of 10K toots would take \url{freespeechextremist.com} just 12 hours, in contrast to \url{social.myfreecams.com} which would take 16 days.
Second,  once this set of toots has been accumulated,  significant manual effort is still required to label them. 
Finding ways to minimise these barriers is therefore vital.

%
%
%
%
%
%
%
%

%
%
%
%
%

\section{ModPair: Decentralised Model Sharing}
\label{sec:modpair}

We next explore collaborative approaches where instances can work together to improve automated toxic content moderation. Namely, we present \emph{ModPair}, a system that facilitates the sharing of pre-trained models between instances. We will show that this can improve the accuracy of local models, with limited overheads for administrators.

\subsection{The Potential of Model Sharing}
\label{sec:modelsharing}

As discussed above, a fundamental challenge is that training the classifier  requires ground-truth labels (\S\ref{sec:annotation_feasibility}). While, for our experiments, we rely on the centralised Perspective API, this is unattractive and infeasible for decentralised instances (due to associated costs and the need to upload toots to centralised third parties). 
This leaves administrators to annotate toots manually, likely as part of their general moderation activities (\eg upon receiving a complaint from a user about a given toot). 
This is both slow and laborious.
We therefore argue that a potential solution is to allow better-resourced instances (in terms of annotations) to share their models with other instances.
In this approach, models trained on one instance are ``gifted'' to another one.

\begin{figure}
     \centering
     \subfloat[][]{
     \includegraphics[width=.45\linewidth]{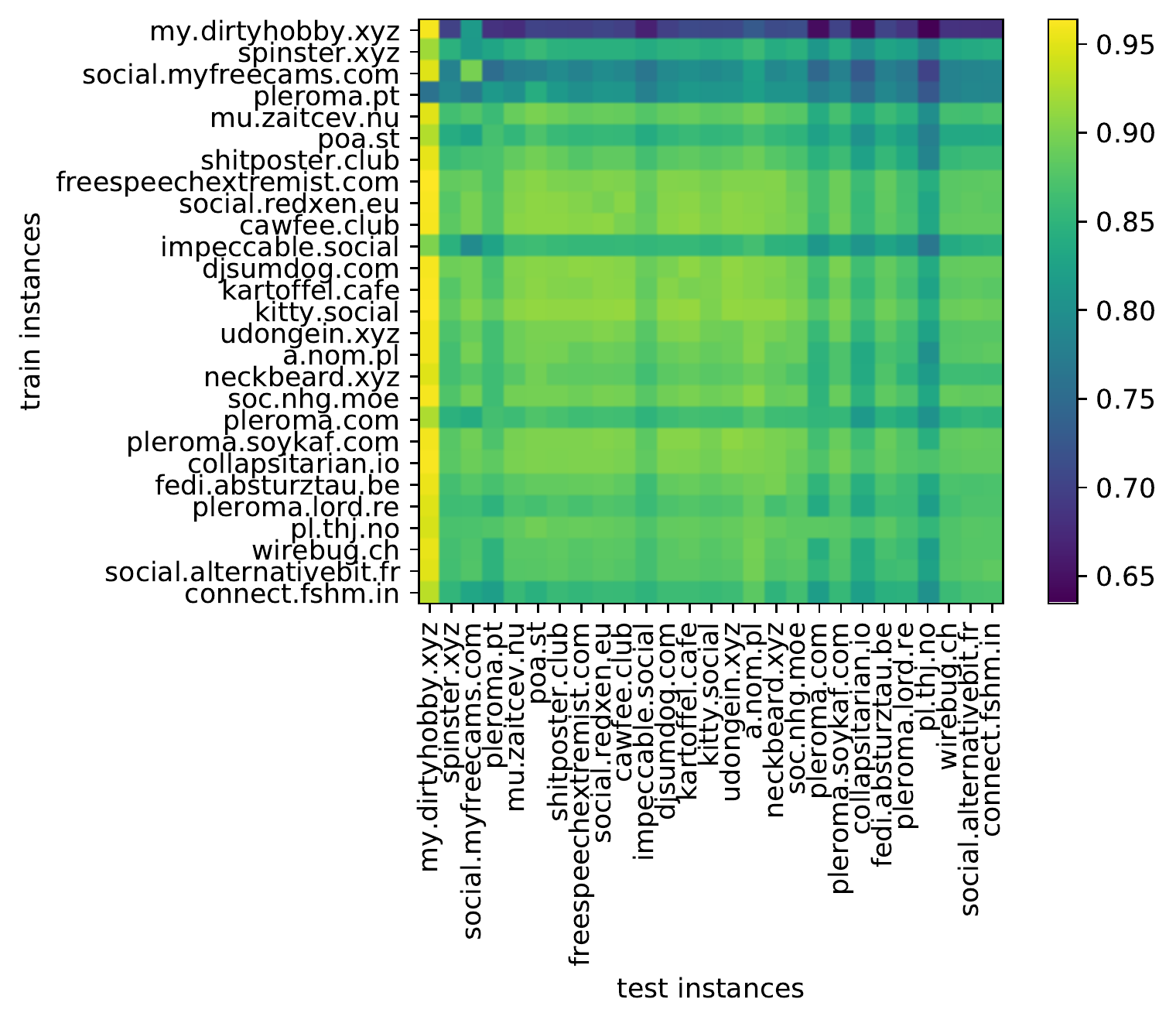} \label{fig:instance_vs_instance_heatmap}}
     \subfloat[][]{\includegraphics[width=.45\linewidth]{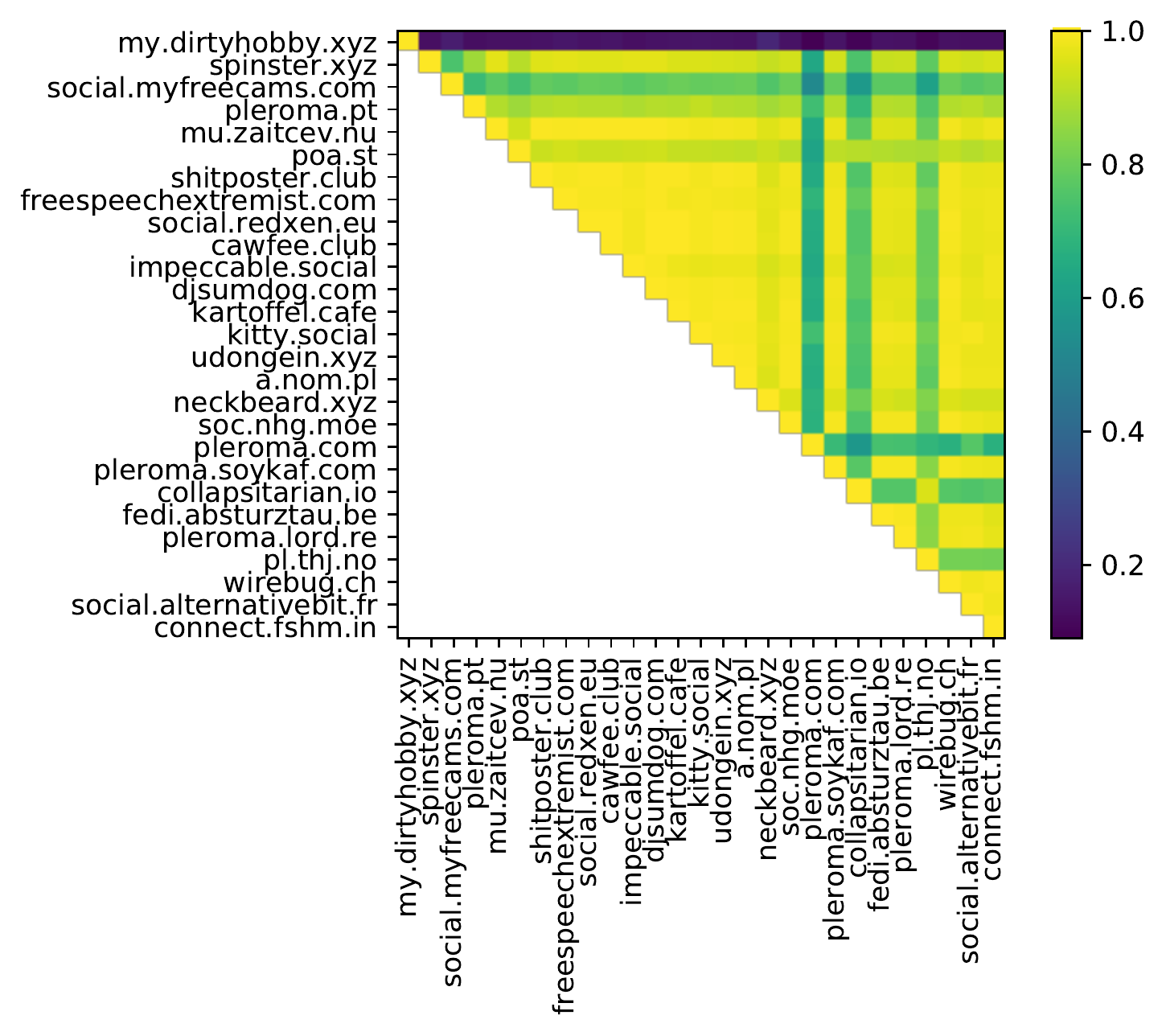}\label{fig:similarityheatmap}}
     \caption{(a) Comparison of models (in terms of macro-F1) trained on one instance (Y-axis) and tested on another instance (X-axis), (b) Comparison of cosine similarity across instances.}
     \label{steady_state}
\end{figure}

To explore the potential of this, following the previous methodology (\S\ref{sec:classifiers}), we \emph{train} a model on each instance using their toots. We then \emph{test} that model on all other instances. This allows us to measure how transferable each model is across instances.
Figure~\ref{fig:instance_vs_instance_heatmap} presents the macro-F1 scores as a heatmap. The Y-axis lists the instances a model has been trained on, and the X-axis lists the instances a model has been tested on. 
We find that performance diverges across the instances. 
The least transferable model is that trained on \url{my.dirtyhobby.xyz}.
Whereas it attains a macro-F1 score of 0.95 when tested on itself, it results in an average of just 0.69 across all other instances, \eg just 0.63 when applied against \url{pl.thj.no}. That said, models trained on some other instances exhibit far greater transferability.
The best performing model is that trained on \url{kitty.social}, which attains an average macro-F1 score of 0.89 across all other instances.
Confirming our observations in \S\ref{sec:performanceacrossinstances}, these trends are largely driven by the types of topics and material shared. For example, we find that 93.4\% of \url{my.dirtyhobby.xyz} toots are identified as sex-related (see Figure~\ref{fig:alltopics}a), whereas the remaining instances contain just 1.4\% of such toots on average. This makes it hard to transfer such models, as their training sets differ wildly from those instances it is applied to. This confirms that decentralised model sharing \emph{can} work effectively, but only in particular cases.

\subsection{ModPair Design}
\label{sec:modpairdesign}

The previous section shows that model sharing \emph{can} work, yet performance is highly variable across instance-pairs due to the differing (linguistic) discourse.
Thus, a clear challenge is identifying which instance models should be shared with other instances. This is not trivial, as it is not possible for instances to scalably inspect all content on each other. 

\pb{ModPair Primer.}
We propose \emph{ModPair}, a system to manage model exchange between instances in the Fediverse. 
ModPair runs on each instance, and executes the following three tasks: 
\one~\emph{Step 1:} It automatically identifies instances suitable for model exchange based on their content;
\two~\emph{Step 2:} It manages model exchange between the instances;
and
\three~\emph{Step 3:} It employs an ensemble to merge results from the top $k$ models, predicted to be most transferable.
ModPair continues to monitor activities across the Fediverse to constantly seek out better matching models. Although we implement ModPair in Pleroma, it is suitable for any other Fediverse platform that follows the same principles (\eg PeerTube, Mastodon).

\pb{ModPair Design.}
\underline{Step 1:} ModPair is first responsible for identifying instances with whom models should be exchanged.
To achieve this, ModPair exploits \emph{content similarity} between instances to predict pairs that should exchange their pre-trained moderation models. 
To achieve this, each instance locally generates a vector with each component corresponding to the \emph{tf-idf} of words from the instance's toots. Specifically, we define this as:

\[ tfidf(t, d) = tf(t, d) . idf(t) \]
where \[ idf(t) = log [ (1 + n) / (1 + df(t)) ] + 1 \]

\noindent and $n$ is the total number of toots in the instance and $df(t)$ is the number of toots in the instance that contain the term $t$.
Instances then exchange these vectors, such that each instance can compute the \emph{cosine similarity} between its own vector and all other vectors. This provides a measure of linguistic closeness between any two instances.
To motivate this design choice, 
Figure~\ref{fig:similarityheatmap} presents the empirical cosine similarity across the instances as a heatmap. 
By comparing this against Figure~\ref{fig:instance_vs_instance_heatmap}, we see that content similarity \emph{does} correlate with model performance, confirming the ability for distance to serve as a predictor for model performance.

\underline{Step 2:} Once similarity has been locally computed, each instance must decide which other instances to download models from.
To do this, ModPair uses a rank threshold, $k$, which stipulates the number of models an instance should retrieve. Specifically, an instance retrieves models from the $k$ other instances that have the smallest cosine similarity (with its own local tf-idf vector).
Note, downloading such models is very lightweight --- using our dataset, we find that the largest model is just 7.7 MB, and the average size is only 3.2 MB per instance.
The reciprocity of this process also helps incentivise participation.
\underline{Step 3:} Upon receipt of the pre-trained model(s), the instances can then use them to perform ensemble-based majority voting to classify future incoming toots.

\pb{ModPair Implementation.}
We have implemented ModPair as part of the Pleroma server software.
In our implementation, each Pleroma instance exposes two new API endpoints to enable the above functionality: 
\one one to publish their tf-idf vector (\texttt{<instance.uri>/api/v1/tfidf});
and \two one to share their local model (\texttt{<instance.uri>/api/v1/model}). 
Any instance requiring moderation models can therefore retrieve the vectors of other instances by communicating with their \texttt{tfidf} endpoint. 
By default, each instance will retrieve updated vectors from all other known instances each week.

\pb{ModPair Scalability.}
By default, ModPair selects the moderation models for an instance by computing the tf-idf cosine similarity with \emph{all} other instances.
Although straightforward, this may result in scalability issues as the number of instances grows (due to the $O(n^2)$ complexity). For example, the 1360 Pleroma instances in our dataset (\S\ref{sec:dataset}) would generate 1.8M transactions alone. Although ModPair rate limits its queries to avoid overwhelming remote instances, this workload is still undesirable. 

To overcome the above challenges, ModPair introduces the optional concept of \emph{pre-sampling} to select a subset of instances to exchange vectors directly with. 
This pre-sampling strategy is driven by the observation that often the best performing models are exchanged among instances that have a high degree of federation.
Thus, on each instance, ModPair ranks all other instances by the number of followers in the target instance. This estimates the amount of social engagement between users on the two instances.
We then pick the top $f$ instances (default 5) with the most shared followers.
Using this, each instance only retrieves the \emph{tf-idf} vectors from these $f$ instances and, therefore, only computes ModPair similarity amongst them. 
As before, we then select the top-3 most similar instances from this set of $f$ options.
Although this potentially misses higher-performing models, this massively reduces the number of retrievals and, consequently, the data exchanged between instances from 1.8M to 6,800 (assuming $f=5$).

\subsection{Evaluation Results}

\pb{Performance of Model Selection.}
To evaluate the correctness of ModPair's selection of models, we calculate its Precision@k (or P@k)~\cite{Craswell2009}, where $k$ is the similarity rank threshold ($k={1,3}$). The precision reflects the proportion of the top-$k$ ModPair predictions that are correct.
In other words, this tests if ModPair selects the best models available.
As an optimal baseline, we compare this against an oracle that pairs each instance with the three other instances whose models achieve the best performance on its local toots (as calculated in Figure~\ref{fig:instance_vs_instance_heatmap}). 
Overall, ModPair achieves average (calculated over all instances) P@1 and P@3 scores of 0.74 and 0.83, respectively. In other words, 74\% of the top-1 and 83\% of the top-3 model selections of ModPair are correct.
This confirms the viability of model sharing, where instances can request models from other similar instance(s) and use them for the moderation of their content. Importantly, this confirms the appropriateness of using lightweight tf-idf vectors to represent instance content.

\begin{figure}
     \centering
     \subfloat[][]{
     \includegraphics[width=.45\linewidth]{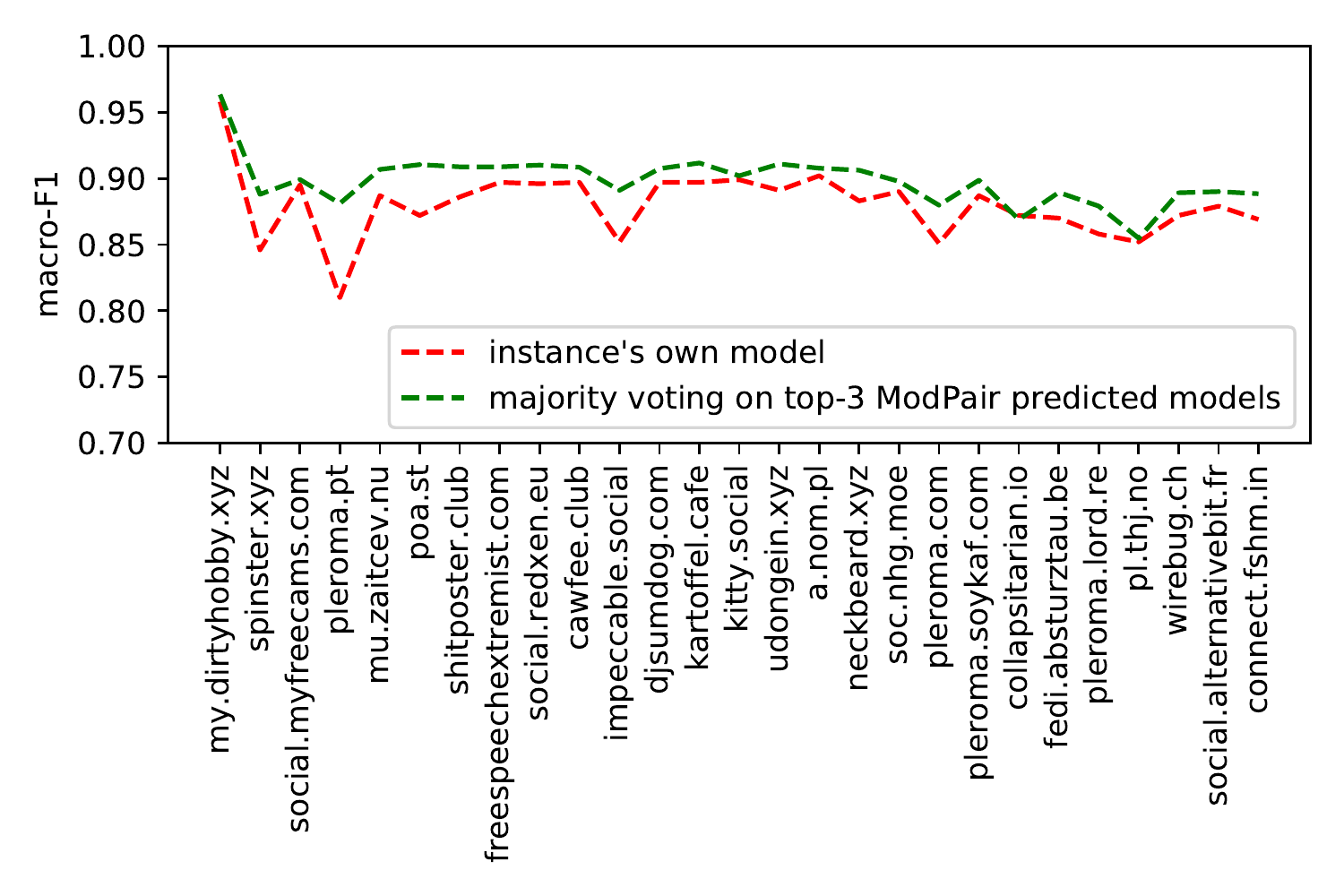} \label{fig:modpairevel}}
     \subfloat[][]{\includegraphics[width=.45\linewidth]{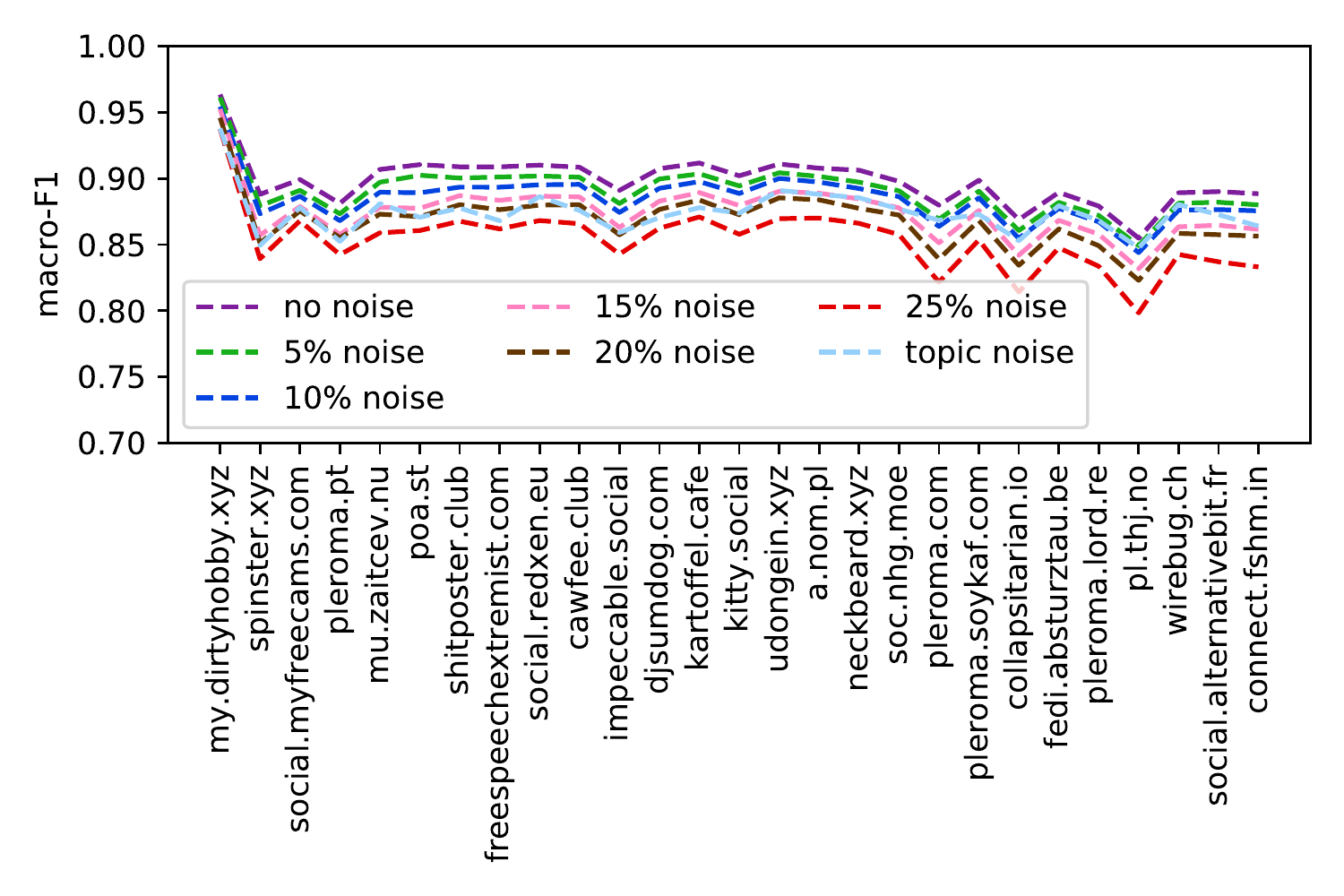}\label{fig:modpairevelnoisy}}
     \caption{Comparison of macro-F1 scores of (a) instance's own model \vs the majority voting on top-3 ModPair predicted models for each instance, (b) majority voting on top-3 ModPair predicted models with varying noise percentage for each instance.}
     \label{steady_state_5}
\end{figure}

\pb{Performance of Model Inference.}
We next test if ModPair improves the outcomes of the model inference. 
To evaluate the impact of ModPair on model performance, for each instance, we apply the top-3 ModPair predicted models (in a majority voting fashion) to its toots. This allows us to calculate the newly attainable macro-F1. We compare this with each instance's own model (\ie locally trained). 
Figure~\ref{fig:modpairevel} presents the results.
We see that consistently ModPair outperforms local training across nearly all instances. 
On average, ModPair increases the macro-F1 by 5.9\% compared to when each instance uses its own model alone (calculated over all instances). The results are also statistically significant as measured using Student's t-test with p $=$ 0.01. Furthermore, ModPair also reduces the number of models (and consequently the number of annotated toots) required to moderate all the instances. For our collection of instances, it reduces the number of models to 10.

\pb{Performance of Model Inference with Noise.}
Recall, we conjecture that some training sets (and models) may contain noise, \eg due to annotation mistakes by administrators (see \S\ref{sec:performanceacrossinstances}). 
We next analyse the inference performance of ModPair in scenarios where individual instance models might be noisy (\eg due to annotation inconsistency). 
To do so, for each instance, we apply the noisy versions (see \S\ref{sec:performanceacrossinstances}) of the top-3 ModPair predicted models (in a majority voting fashion) to its toots and  calculate the newly attainable macro-F1. Figure~\ref{fig:modpairevelnoisy} presents the results. We observe that even with 25\% noise induced into individual models, this
results in limited performance degradation of just an  average of 4.4\% macro-F1 reduction (calculated over all instances).
This is significantly lower than the degradation observed earlier when only using a single locally trained model (which obtained -11.9\%).
For the noisy models where we flipped the labels of toxic toots belonging to a particular topic, we observe the average degradation in performance of only 2.2\%. This occurs because instances are being paired with others that have similar linguistic features. Therefore, this has little impact on overall performance because the paired instances are flipping similar topics.
This further confirms the applicability of ModPair.

%
%

\begin{figure}[t]
\centering
\includegraphics[width=0.7\columnwidth]{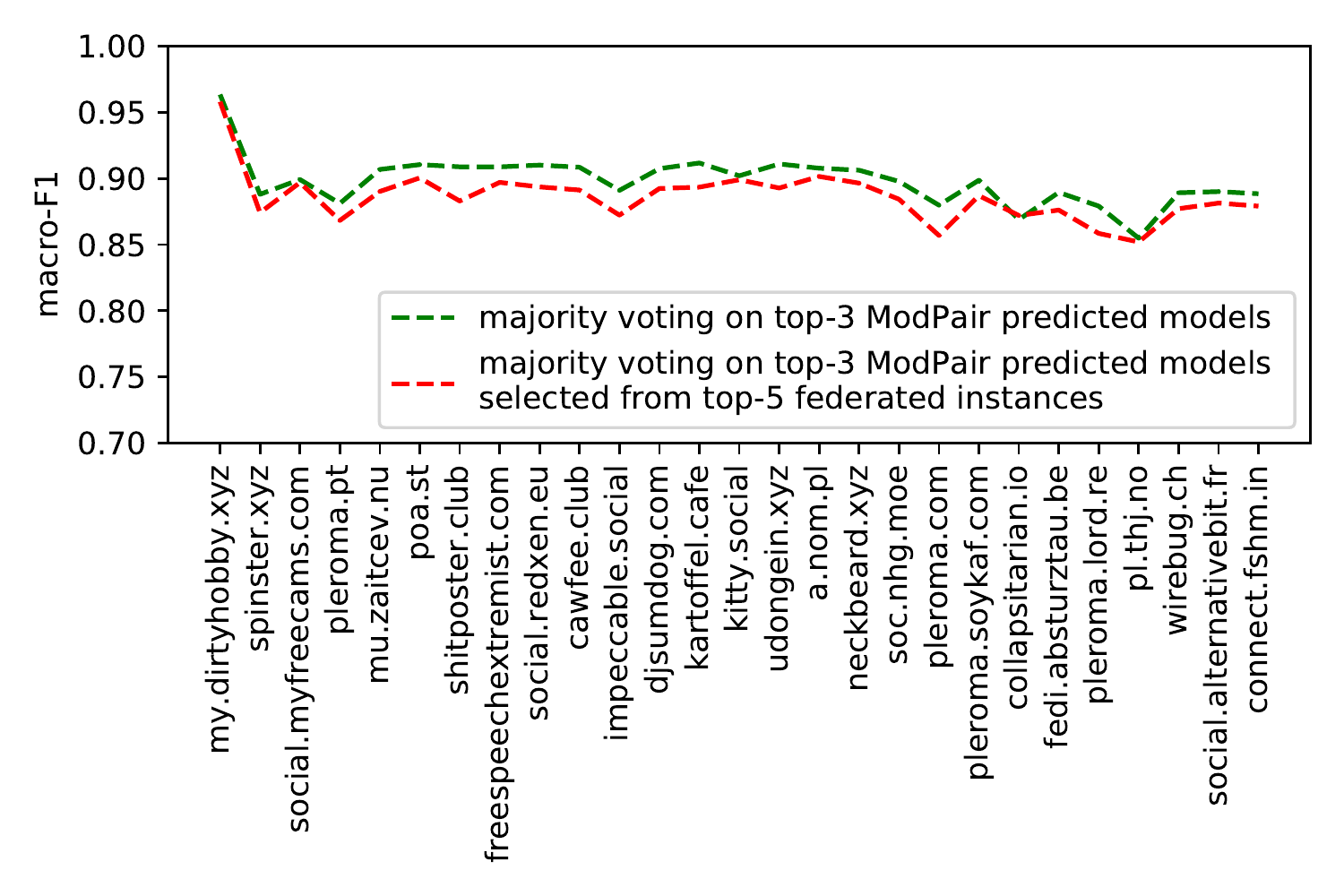} 
\caption{Comparison of macro-F1 scores of the majority voting on top-3 ModPair predicted models selected from the pool of all instances \vs selected from the pool of 5 federated instances.}
\label{fig:modpairevelfederated}
\end{figure}

\pb{ModPair Scalability.}
The prior results are based on $O(n^2)$ model comparisons. 
Therefore we next repeat our experiments, using our scalable pre-sampling strategy that selects a subset of $f$ instances to exchange $td-idf$ vectors with. 
Figure~\ref{fig:modpairevelfederated} compares the performance against the prior approach of retrieving all vectors. 
Confirming our intuition, we only observe a slight degradation (1.1\%) in the average performance compared to when models are predicted from the pool of all instances.
This occurs because instances often federate with others who discuss similar topics, therefore increasing the probability that their models may transfer well. 
Despite the small degradation in performance, we also emphasise that the average performance is still 4.7\% better than when each instance uses its own model (\S\ref{sec:performanceacrossinstances}).

\subsection{Discussion}
Model sharing via ModPair allows instances to exchange models from other similar instances. This raises several issues worthy of discussion.

\pb{Privacy.}  ModPair enables instances to share the trained models (specifically share weights and biases of the trained models). In terms of privacy, this is better than sharing the raw data itself, preserving the privacy of toot text, as well as the annotations given by other administrators. Further, sharing of tf-idf vectors is privacy-preserving as they consider the words independently, hence it is not possible to reconstruct the original data.
That said, studies show~\cite{lundberg2017unified, murdoch2018beyond}, through reverse-engineering techniques, some important words or phrases used during the training can be identified. Although not explored here, we also note that there are protocols for computing cosine similarity and model sharing with privacy guarantees, which could be integrated with ModPair~\cite{gheid2015efficient, mo2021ppfl}.

\pb{Security in Adversarial Contexts.} ModPair must also ensure it is not exploited by malicious parties. Most obviously, malicious instances could purposefully provide invalid models or tf-idf vectors to undermine other administrators. To mitigate this, ModPair's pre-sampling leverages follower relationships to better identify trusted similar instances. Nevertheless, this is a topic ripe for further exploration. We envisage prior work on peer-to-peer reputational models will be extremely useful here~\cite{kamvar2003eigentrust}.

\pb{Fairness \& Bias.} There are chances that a model trained on a large instance with a large number of toots could be paired with a relatively small instance for model transfers.
If such models are used widely, this could allow a small set of instances to bias the overall moderation process. Furthermore, this may ``drown out'' models from more fringe instances that might be useful for identifying specific forms of toxic content.
While this is not necessarily a problem, as ModPair only pairs instances with similar interests, it may be necessary to adjust voting weights to prevent bias. We also plan to explore alternate ways of computing similarity that consider other factors such as the number of toots on each instance.

\section{Related Work}

\pb{Decentralised Social Network Measurements.}
Prior work has extensively studied social networks with respect to their structure~\cite{ahn2007analysis,leskovec2008planetary,cheng2008statistics,traud2012social,magno2012new,myers2014information,manikonda2014analyzing} and evolution~\cite{viswanath2009evolution,kumar2010structure,gonzalez2013google+}.
The overwhelming majority of studies have been conducted on \emph{centralised} social networks such as Twitter~\cite{ribeiro2018characterizing,mathew2019spread}, with only a handful of papers focusing on \emph{decentralised} social networks. 
In~\cite{datta2010decentralized}, Datta \etal explore various motivations for decentralised  social networking. Schwittmann \etal~\cite{schwittmann2013sonet} analyze the security and privacy of decentralised social networks.
Bielenberg et al.~\cite{bielenberg2012growth} shed light on the evolution of one of the first decentralized social networks, Diaspora, discussing its growth in terms of the number of users, and the topology of Diaspora's interconnected servers. 
Raman~\etal~\cite{raman2019challenges} identify key challenges in the decentralised web, mainly related to network factors~\cite{kaune2009modelling}.
Finally, ~\cite{zignani2018follow} authors collect data from Mastodon and explore several features such as the relationship network, placement of instances, and content warnings.

To the best of our knowledge, we are the first to measure the spread of toxicity on a decentralised social network. 
The independent and interconnected nature of instances makes this particularly different to prior studies of centralised platforms.
Closest related to our work is \cite{zignani2019mastodon}, which gathered a dataset of content warnings from Mastodon (another DW platform). However, as shown in \S\ref{sec:performanceacrossinstances}, we find that these are unsuitable for training classification models.

\pb{Toxic Content Classification.}
There have been a variety of works that attempt to build classifiers for the automatic identification of toxic content~~\cite{van2018challenges, vidgen2019challenges, vidgen2020directions}.
These include fundamental work on creating formal toxic content definitions~\cite{sheth2021defining} and compiling vital datasets~\cite{arhin2021ground,de2018hate,guest2021expert}.
Of particular interest are prior attempts to build text classifiers to automatically flag toxic material.

Deep learning has been widely used to classify toxic posts~\cite{badjatiya2017deep, rui2020deephate}.
The effectiveness of these mechanisms has been extensively evaluated~\cite{chandrasekharan2017you, jhaver2018online, matakos2017measuring, vidgen2019challenges}, alongside the design of techniques to circumvent such tools~\cite{gerrard2018beyond}.
Badjatiya \etal~\cite{badjatiya2017deep} were amongst the first to show the benefits of training models to identify hate speech (using 16K annotated tweets). 
Rizos \etal~\cite{rizos2019augment} evaluate the performance data augmentation techniques applied on conventional and recurrent neural networks trained on Yahoo~\cite{warner2012detecting} and Twitter~\cite{davidson2017automated}. Wulczyn \etal~\cite{wulczyn2017ex} similarly use 100K user labelled comments from Wikipedia to identify the nature of online personal attacks.
Others have shown that more traditional machine learning models such as n-grams and Logistic Regression~\cite{waseem2016hateful} SVMs~\cite{sood2012automatic}, Na\"ive Bayes, decision trees and random forests~\cite{davidson2017automated} also produce good results. 
There has also been work identifying controversial posts by using non-text features, such as user profile metadata~\cite{weimer2007automatically, vicenc2008statistical, rios2012dissimilarity}.

Related to our work are recent efforts in the area of transfer learning~\cite{devlin2018bert, brown2020language}. 
In terms of toxic content identification, these exploit pre-trained language models such as BERT, and perform fine-tuning to specialise the model~\cite{wei2021offensive, mozafari2019bert}.
This subsequently `transfers' semantic understanding from the general language model to the task in-hand. 
Although similar in concept, we take a very different approach to transferring knowledge between instances. 
Specifically, ModPair relies on model exchange and ensemble voting. 
This is because transfer fine-tuning~\cite{benchmarking2020train} and even inference~\cite{benchmarking2020infer} are substantially more costly than lightweight models such as Logistic Regression (considering the limited capacity of most Pleroma instances). In our future work, we intend to explore the potential of using lightweight pre-trained models such as LadaBERT~\cite{mao2020ladabert}.

\pb{Moderation in Social Networks.}
There have been a number of studies that have applied the above classification models to measure toxic activities within social networks, including
Twitter~\cite{ribeiro2018characterizing,burnap2015cyber,waseem2016hateful}, 
Reddit~\cite{chandrasekharan2017you,mohan2017impact,almerekhi2020investigating}, 
4Chan~\cite{bernstein20114chan,papasavva2020raiders},
as well as various fringe forums~\cite{de2018hate}.
There have also been studies that focus on specific domains of toxicity, including anti-Semitism~\cite{warner2012detecting}, %
cyberbullying~\cite{chatzakou2019detecting},
white supremacy~\cite{de2018hate}, misogyny~\cite{guest2021expert}, and Islamaphobia~\cite{magdy2016isisisnotislam}. More recently, Anaobi~\etal~\cite{hassan2021exploring} highlighted the content moderation challenges for the instance administrators in decentralised web. We contribute to this wider space, by focusing on characterising the spread of toxicity in a novel platform: Pleroma. 
Beyond these prior works, we also show how such content spreads across independently operated instances (a unique feature of the Fediverse).
As part of this, we identify and quantify hitherto unknown challenges for moderation that result from the Fediverse's decentralised architecture.
Finally, we propose and evaluate a solution, ModPair, that allows instances to automatically exchange models to support scalable decentralised moderation.

\section{Conclusion \& Future Work}
\label{sec:conclusion}

This paper has examined toxicity in Pleroma, a popular Decentralised Web (DW) social platform. 
We have characterised the spread of toxicity on the platform, confirming that the federation process allows toxic content to spread between instances. 
We have further explored the challenges of moderating this process by building per-instance models. We found that whilst this can be effective, it comes with a heavy burden on administrators who must annotate toots. 
To reduce this burden and enable collaboration amongst instances, we presented ModPair, a system for pairing instances that can share pre-trained models. We showed that by exchanging models with just a small set of other instances, administrators can effectively collaborate to improve each others' detection accuracy.
Further, we have shown how ModPair can be scaled-up by using pre-sampling to dynamically select remote instances that are likely to have a close linguistic similarity.
Our work contributes to the wider debate on online toxic behaviour, as well as offers tools that can support the growth of platforms with expanding importance within the Fediverse.

As part of our future work, we plan to expand our analysis to other DW platforms and investigate other behavioural attributes, including the time-variance of tf-idf vectors (\ie as instances generate toots, their tf-idf representation will change with time). We also intend to experiment with alternative features (\eg semantically-rich sentence embeddings~\cite{reimers2019sentence}) as well as classification approaches (\eg allowing more resource-capable instances to train and share LSTM~\cite{hochreiter1997long} or BERT-based models~\cite{devlin2018bert}). 
As part of this, we are curious to investigate the feasibility of using other types of Federated Learning~\cite{mcmahan2016federated, mo2021ppfl} too.
Finally, we are keen to better understand the discrepancies between the annotations performed by different administrators (\eg in terms of how they locally define toxicity). Through this, we hope to gain further evaluative insight into how ModPair can be useful to administrators.

\begin{acks}
 This work is supported by EPSRC grants
 SODESTREAM (EP/S033564/1),
AP4L (EP/W032473/1), 
EPSRC REPHRAIN  ``Moderation in Decentralised Social Networks'' (DSNmod),
and
EU Horizon grant agreements No 830927 (Concordia) and No 101016509 (Charity).
\end{acks}

\bibliographystyle{ACM-Reference-Format}
\input{no_comments.bbl}

\appendix

\section{Additional Experiments}
\label{sec:additionalexperiments}

\begin{figure}[h]
     \centering
     \subfloat[][using toxicity threshold of 0.8.]{
     \includegraphics[width=.45\linewidth]{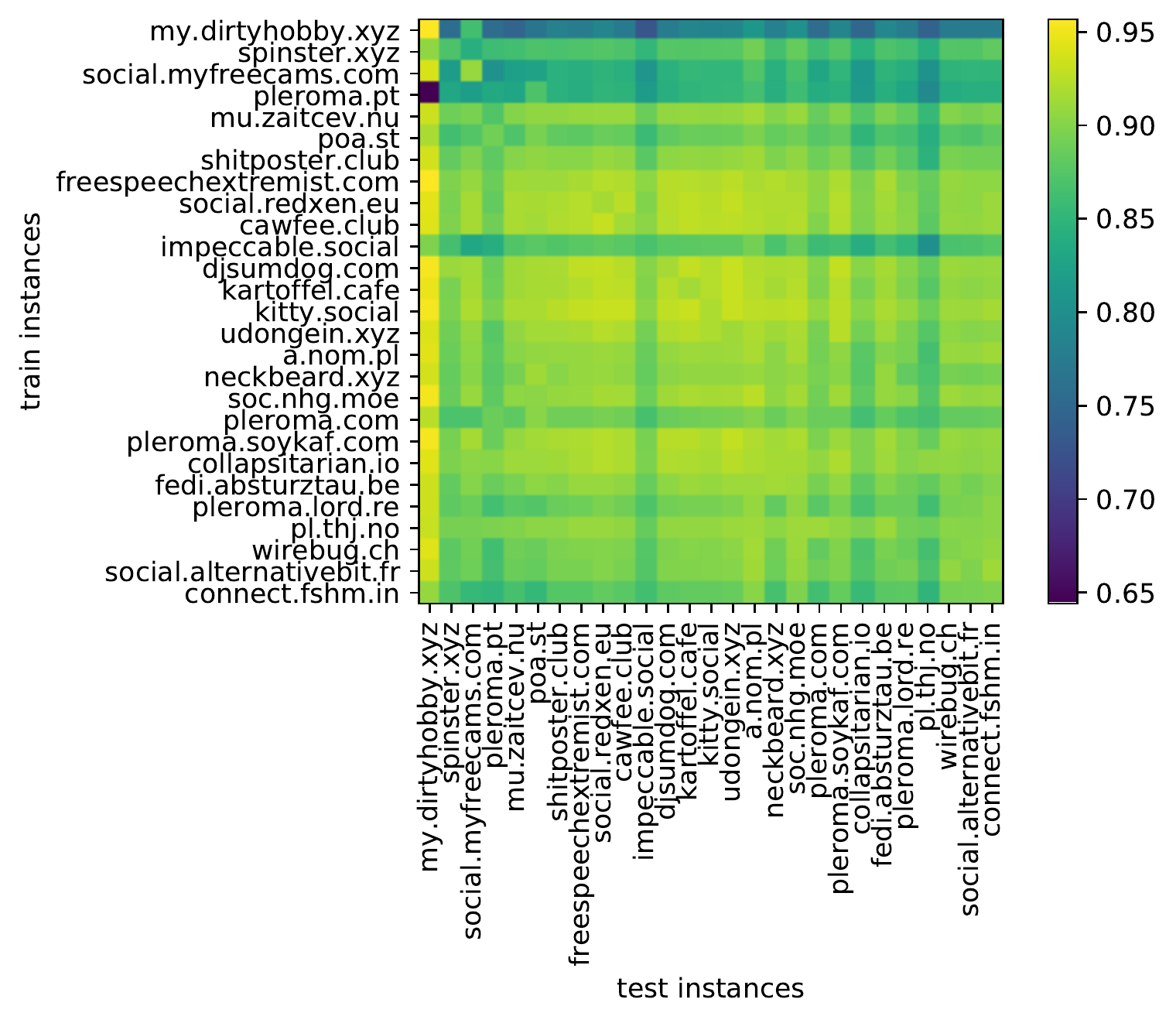} \label{fig:heatmap_tox_0point8}}
     \subfloat[][using SVM classifier.]{\includegraphics[width=.45\linewidth]{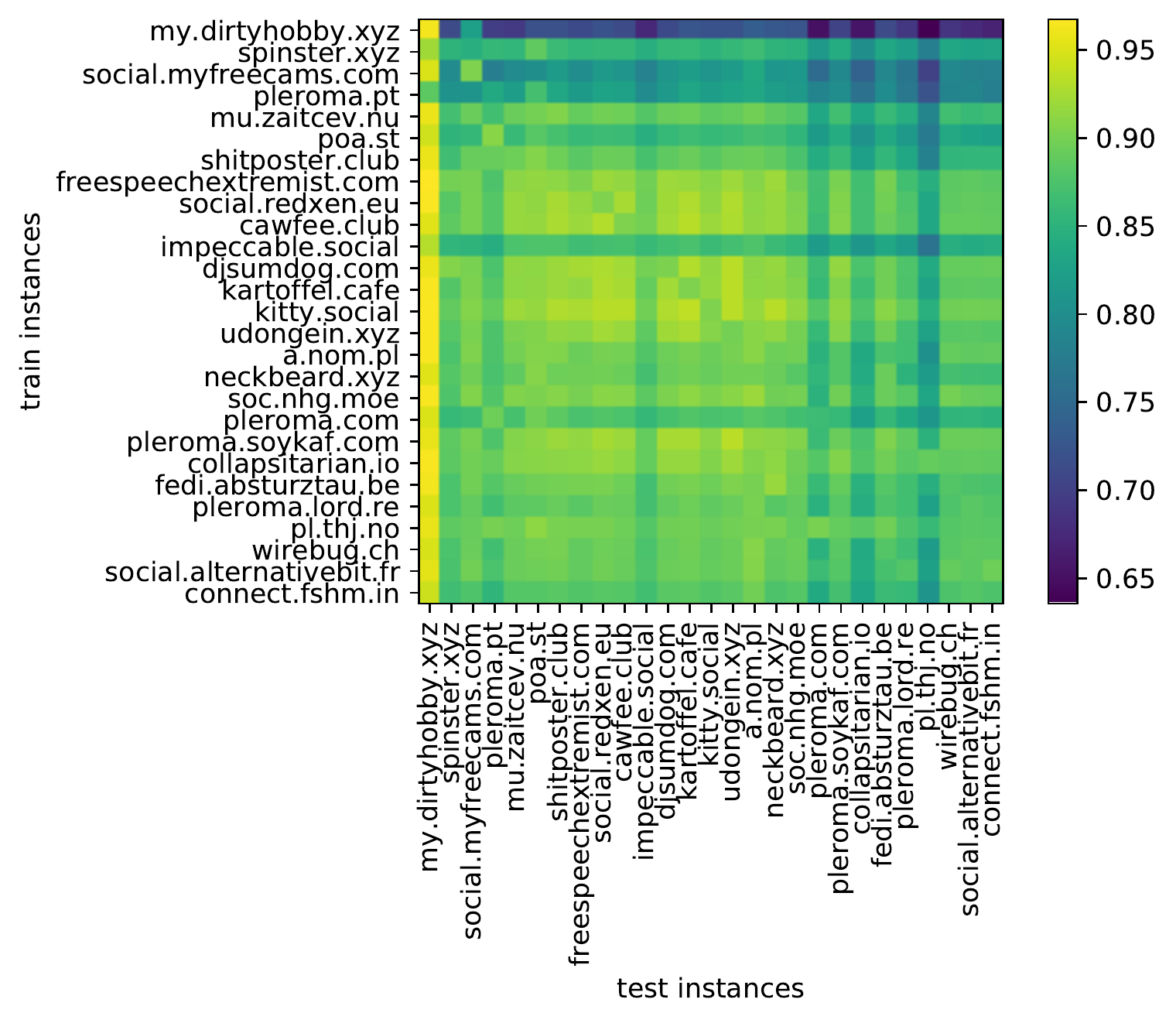}\label{fig:heatmap_SVM}}
     \caption{Comparison of models (in terms of macro-F1) trained on one instance (Y-axis) and tested on another instance (X-axis).}
     \label{steady_state_4}
\end{figure}

In addition to experimenting with a toxicity threshold of 0.5, we also conducted experiments with a much stricter threshold 0.8. In this case, we considered a toot toxic if its toxicity score is $>$ 0.8 (and vice-versa). We performed $N^2$ experiments as shown in Figure~\ref{fig:heatmap_tox_0point8} and the results were similar to that obtained in case of 0.5 threshold.

Furthermore, we also conducted experiments with a SVM classifier and, as shown in Figure~\ref{fig:heatmap_SVM}, results are comparable to Logistic Regression. 
These additional experiments further emphasize that our methodology and approach can be reused with other labeling schemes and classifiers.

\begin{figure}[h]
     \centering
     \subfloat[][With toxicity threshold of 0.8.]{
     \includegraphics[width=.32\linewidth]{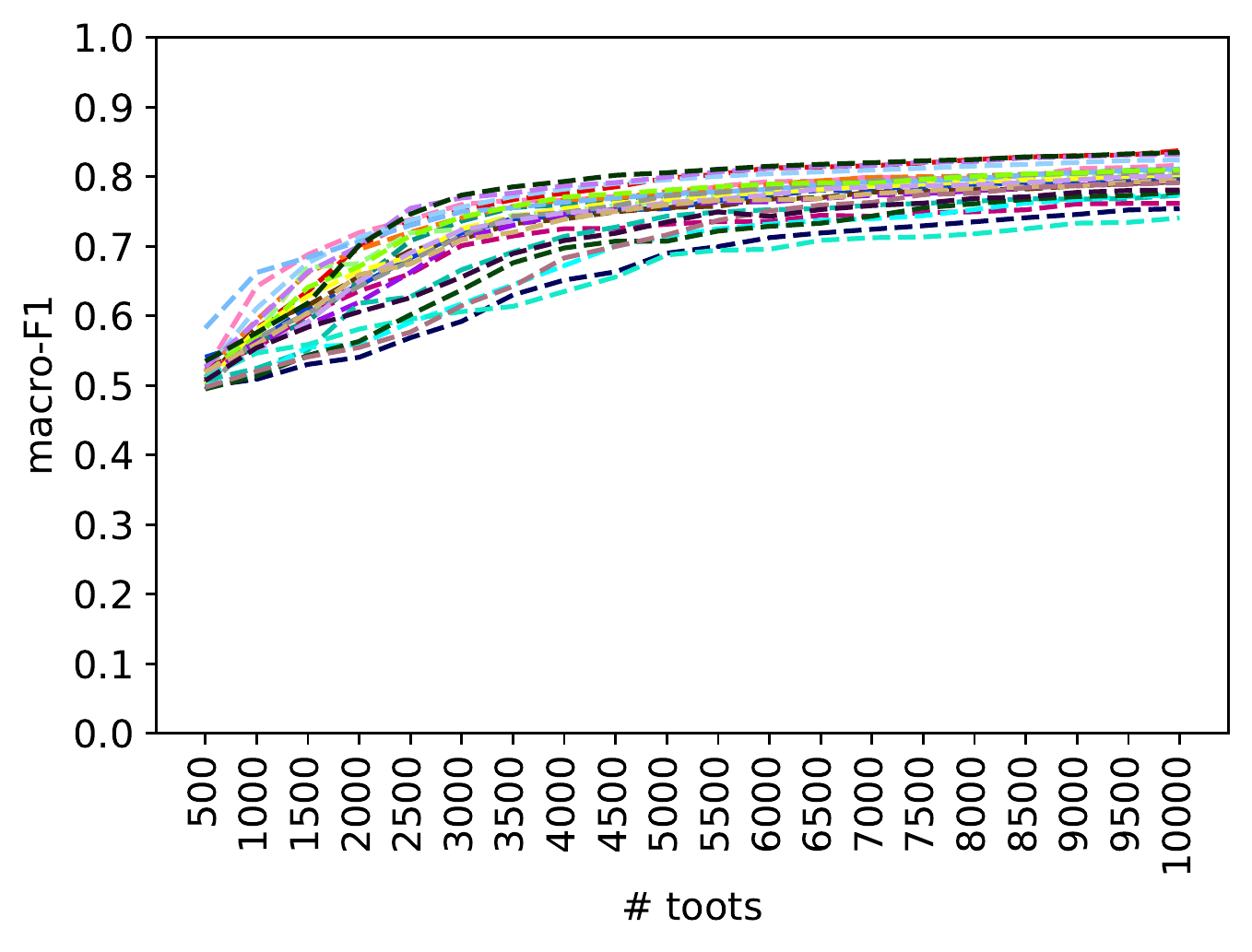} \label{fig:classifierNrandom_tox0point8}}
     \subfloat[][With SVM classifier.]{\includegraphics[width=.58\linewidth]{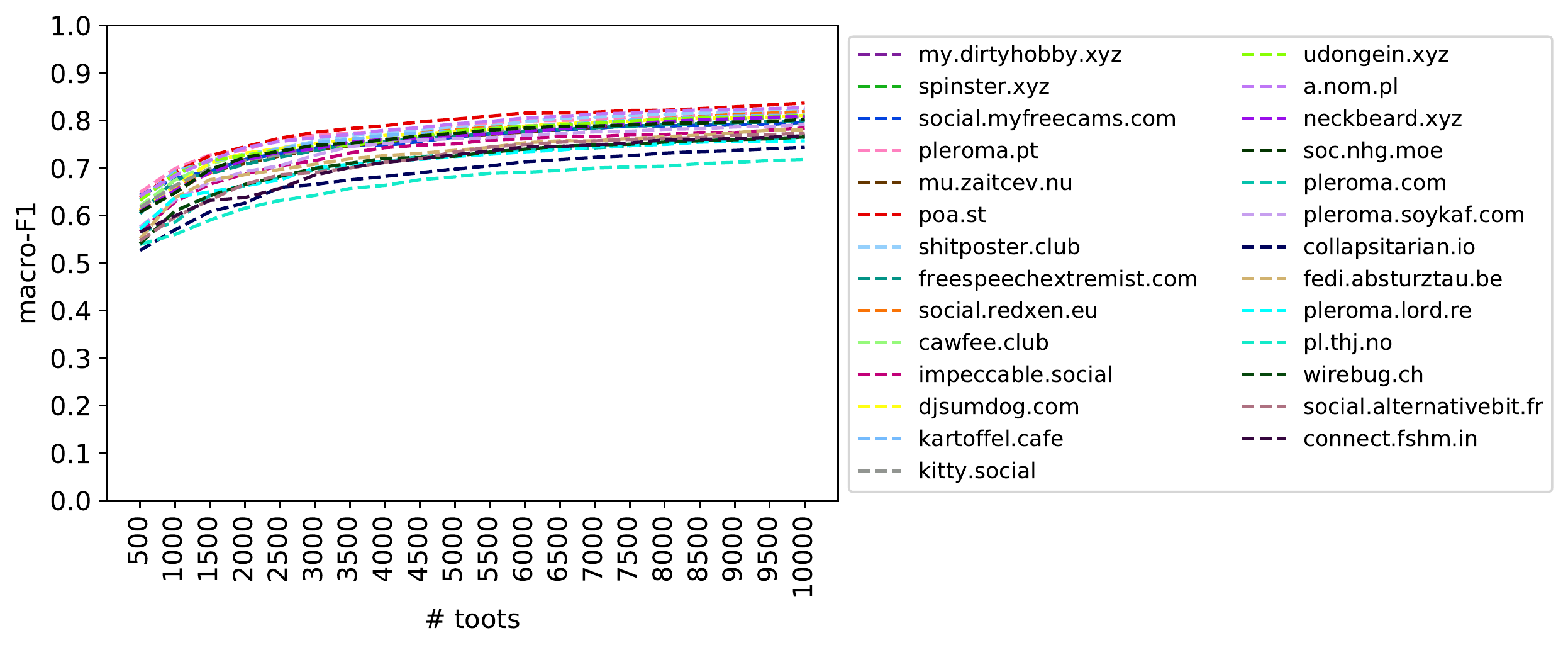}\label{fig:classifierNrandom_SVM}}
     \caption{Macro-F1 scores of classifiers trained on random $n$ toots.}
     \label{steady_state_6}
\end{figure}

\end{document}